\documentclass{iopart}
\usepackage[utf8]{inputenc} %
\usepackage[colorlinks=true, citecolor=blue, filecolor=black,linkcolor=blue, urlcolor=blue, hypertexnames=false]{hyperref}
\usepackage{color}

\usepackage{array,multirow}
\usepackage{enumitem}

\usepackage{rotating}
\usepackage{multirow}

\expandafter\let\csname equation*\endcsname\relax
\expandafter\let\csname endequation*\endcsname\relax
\usepackage{amsmath}
\usepackage{amssymb}
\usepackage{amsmath}
\usepackage{amsfonts}
\usepackage{amsthm}
\usepackage{iopams}
\usepackage{subdepth}

\usepackage{graphicx}
\usepackage{geometry}
\usepackage{subcaption}
\usepackage[ruled]{algorithm2e}
\theoremstyle{definition}
\newtheorem{theorem}{Theorem}[section]
\newtheorem{proposition}[theorem]{Proposition}

\newtheorem{assumption}[theorem]{Assumption}

\theoremstyle{definition}

\newtheorem{example}[theorem]{Example}

\theoremstyle{remark}
\newtheorem{remark}[theorem]{Remark}
\newcommand{\R}{\mathbb{R}}
\newcommand{\E}{\mathbb{E}}
\renewcommand{\P}{\mathbb{P}}

\DeclareMathOperator{\trace}{trace}

\renewcommand{\d}{\mathrm{d}}
\newcommand{\Dkl}{D_\mathrm{KL}}

\usepackage[textwidth=1.05in]{todonotes}

\renewcommand{\ldots}{{.\hss.\hss.}}
\newcommand{\tx}[1]{\mathrm{#1}}

\begin{document}

\title[Data-Free Likelihood-Informed Dimension Reduction]{Data-Free Likelihood-Informed Dimension Reduction of Bayesian Inverse Problems}

\author{Tiangang Cui$^{1}$, Olivier Zahm$^{2}$}
\address{$^1$School of Mathematics, Monash University, VIC 3800, Australia}
\address{$^2$Univ. Grenoble Alpes, Inria, CNRS, Grenoble INP, LJK, 38000 Grenoble, France}
\ead{\mailto{tiangang.cui@monash.edu}, \mailto{olivier.zahm@inria.fr}}

\begin{abstract}
Identifying a low-dimensional informed parameter subspace offers a viable path to alleviating the dimensionality challenge in the sampled-based solution to large-scale Bayesian inverse problems. This paper introduces a novel gradient-based dimension reduction method in which the informed subspace does not depend on the data. This permits an online-offline computational strategy where the expensive low-dimensional structure of the problem is detected in an offline phase, meaning before observing the data. This strategy is particularly relevant for multiple inversion problems as the same informed subspace can be reused. The proposed approach allows controlling the approximation error (in expectation over the data) of the posterior distribution. We also present sampling strategies that exploit the informed subspace to draw efficiently samples from the exact posterior distribution. The method is successfully illustrated on two numerical examples: a PDE-based inverse problem with a Gaussian process prior and a tomography problem with Poisson data and a Besov-$\mathcal{B}^2_{11}$ prior.
\end{abstract}

\section{Introduction}

The Bayesian approach to inverse problems builds a probabilistic representation of the parameter of interest conditioned on observed data. 
Denoting the parameter and data by $x\in\R^d$ and $y\in\R^m$, respectively, the solution to the inverse problem is encapsulated in the posterior distribution, which has the probability density function (pdf)
\begin{equation}\label{eq:BayesianModel}
 \pi_\text{pos}^y(x)
 = \frac{1}{\pi_\text{data}(y)} \mathcal{L}^y(x)  \pi_\text{pr}(x) ,
\end{equation}
where $\pi_\text{pr}(x) $ denotes the prior density, $\mathcal{L}^y(x)$ is the likelihood function, and $\pi_\text{data}(y)$ is the marginal density of the data $y$ that can be expressed as
\begin{equation}\label{eq:PiData}
 \pi_\text{data}(y)  = \int_{\R^d} \mathcal{L}^y(x)  \pi_\text{pr}(x) \, \d x.
\end{equation}
This way, one can encode the posterior into summary statistics, for example, moments, quantiles, or probabilities of some events of interest \cite{kaipio2006statistical,IP:Stuart_2010,tarantola2005inverse}, to provide parameter inference and associated uncertainty quantification. 
In practice, computing these summary statistics requires dedicated methods to efficiently characterize the posterior distribution. 
Markov chain Monte Carlo (MCMC) methods \cite{MCMC:BGJM_2011,liu2001monte}, originating with the Metropolis-Hastings algorithm \cite{MCMC:Hastings_1970,MCMC:Metropolis_etal_1953}, and sequential Monte Carlo methods \cite{MCMC:SMC_2006} have been developed as workhorses in this context.
However, many inverse problems have high-dimensional or infinite-dimensional parameter space, which present a significant hurdle to the applicability of MCMC, SMC, and other related sampling methods in general.

The efficiency of these sampling methods, measured by the required number of posterior density evaluations, may deteriorate with the dimension of the parameter space, see \cite{MCMC:RGG_1997,roberts2001optimal} and references therein. Even with the rather strong log-concave assumption, start-of-the-art MCMC methods can still be sensitive to the dimension of the problem, see for instance \cite{dalalyan2017theoretical, durmus2019high,dwivedi2019log}.

One promising way to alleviating the challenge of dimensionality is to exploit the effectively low-dimensional structures of the posterior distribution.
Such low-dimensional structures can be used to construct certified low-dimensional approximations of the posterior distribution \cite{spantini2015optimal,zahm2018certified} and efficient MCMC proposals that are robust in the parameter dimension \cite{beskos2017geometric,beskos2018multilevel,cui2016dimension,cui2019multilevel,lan2019adaptive}. 
There exists several ways to detect low-dimensional structures.
A widely accepted method is to utilize the regularity of the prior, in which the dominant eigenvectors of the prior covariance operator \cite{marzouk2009dimensionality} can be used to define such a low-dimensional subspace. This prior-based dimension reduction also plays a key role in the analysis of high-dimensional integration methods such as \cite{graham2015quasi,graham2011quasi}. 
In addition to the prior regularity, the limited accuracy of the observations and the ill-posed nature of the forward model often allow one to express the posterior as a low-dimensional update from the prior. 
Methods such as the active subspace (AS) \cite{constantine2016accelerating,cortesi2020forward} and the likelihood-informed subspace (LIS) \cite{cui2014likelihood,cui2021unified,zahm2018certified} utilize gradients of the forward model and/or of the likelihood function in order to better identify the low-dimensional structure of the problem.
We refer to \cite{cui2021unified,zahm2018certified} for an overview and a comparison of the existing methods.

The success of AS and LIS relies on the computation of the gradient or the Hessian of the log-likelihood function. Since the likelihood function depends on the observed data, the resulting subspaces need to be reconstructed each time a new data set is observed. This can add a significant computational burden to the solution of inverse problems. 
In this paper, we present a new \emph{data-free} strategy for constructing the informed subspace in which the computationally costly subspace construction can be performed in an \emph{offline phase}, meaning before observing any data sets.
In the subsequent \emph{online phase}, the data set is observed and the precomputed informed subspace is utilized to accelerate the inversion process.
This computational strategy is particularly relevant for real-time systems such as medical imaging where multiple inversions are needed.

The rest of the paper is organized as follows. To begin, we introduce the problem setting in Section \ref{sec2}. In Section \ref{sec3}, we present a new data-free likelihood-informed approach to construct the subspace. Denoting the Fisher information matrix of the likelihood function by $\mathcal{I}(x)=\int (\nabla \log\mathcal{L}^y(x))(\nabla \log\mathcal{L}^y(x))^\top \mathcal{L}^y (x)\d y$, this approach defines the informed subspace as the rank-$r$ dominant eigenspace of the matrix
\begin{equation}\label{eq:h1}
 H = \int \mathcal{I}(x) \, \pi_\text{pr}(x) \d x,
\end{equation}
with $r \ll d$. This definition makes no particular assumption on the likelihood function, so it can be applied to a wide range of measurement processes, e.g., Gaussian likelihood and Poisson likelihood. It also does not involve any particular data set $y$, and hence can be constructed offline. 
Given the informed subspace, we approximate the posterior density $\pi_\text{pos}^y(x)$ by
\begin{equation}\label{eq:a1}
\widetilde\pi_\text{pos}^y(x) = \widetilde\pi_\text{pos}^y(x_r) \pi_\text{pr}(x_\perp|x_r),
\end{equation}
where $x_r$ and $x_\perp$ denote respectively the informed and the non-informed components of $x$. 
We prove that the expected Kullback-Leibler (KL) divergence of the full posterior from its approximation is bounded as
\begin{equation}\label{eq:e1}
 \E[\Dkl(\pi_\text{pos}^Y||\widetilde\pi_\text{pos}^Y) ] \leq \kappa \sum_{i>r} \lambda_i(H) ,
\end{equation}
where the expectation is taken over the data $Y\sim\pi_\text{data}(y)$, $\kappa$ being the subspace Poincar\'e constant of the prior \cite{zahm2018gradient,zahm2018certified} and $\lambda_i(H)$ the $i$-th largest eigenvalue of $H$. 
This way, a problem with a fast decay in the spectrum of $H$ yields an accurate low-dimensional posterior approximation in expectation over the data.

In Section \ref{sec4}, we restrict the analysis to Gaussian likelihood. In this case, we show that the vector-valued extension \cite{zahm2018gradient} of the AS method \cite{constantine2014active}, which reduces  parameter dimensions via approximating forward models, also leads to the same data-free informed subspace as that obtained using \eqref{eq:h1}. We can further show that, although the likelihood-informed approach and AS employ different approximations to the posterior density, the resulting approximations share the same structure as shown in \eqref{eq:a1} and follow the same error bound as in \eqref{eq:e1}.

As suggested by \eqref{eq:a1}, the factorized form of the approximate posterior densities allows for dimension-robust sampling. One can explore the low-dimensional intractable parameter reduced posterior $\widetilde\pi_\text{pos}^y(x_r)$ using methods such as MCMC, followed by direct sampling of the high-dimensional but tractable conditional prior $\pi_\text{pr}(x_\perp|x_r)$. This strategy has been previously investigated, see \cite{cui2014likelihood,zahm2018certified} and references therein. We provide a brief summary to this existing sampling strategy in Section \ref{sec5}. 
Despite the accelerated sampling offered by the informed subspace, the resulting inference results are subject to the dimension truncation error that is bounded in \eqref{eq:e1}. 
In Section \ref{sec6}, by integrating the pseudo-marginal approach \cite{andrieu2009pseudo,andrieu2015convergence} and the surrogate transition approach \cite{christen2005markov,liu2001monte,liu1998sequential} into the abovementioned sampling strategy, we present new \emph{exact inference} algorithms that can enjoy the same subspace acceleration while target on the full posterior. 
Our exact inference algorithms only require minor modifications to the sampling strategy of \cite{cui2014likelihood,zahm2018certified}.

While our dimension reduction method readily apply for Gaussian priors, its application to non-Gaussian priors might not be straightforward.
In Section \ref{sec_besov}, we show how to use the propose method for problems with Besov priors \cite{dashti2012besov,kolehmainen2012sparsity,saksman2009discretization} which are commonly used in image reconstruction problems.

We demonstrate the accuracy of the proposed data-free LIS and the efficiency of new sampling strategies on a range of  problems. These include the identification of the diffusion coefficient of a two-dimensional elliptic partial differential equation (PDE) with a Gaussian prior in Section \ref{sec:numerics_elliptic} and Positron emission tomography (PET) with Poisson data and a Besov prior in Section \ref{sec:poisson_numerics}.

 \section{Problem setting}\label{sec2}

For high-dimensional ill-posed inverse problems, the data are often informative only along a few directions in the parameter space. 
To detect and exploit this low-dimensional structure, we introduce a projector $P_r\in\R^{d\times d}$ of rank $r\ll d$ such that $\text{Im}(P_r)$ is the informed subspace and $\text{Ker}(P_r)$ the non-informed one. This splits the parameter space as
$$
 \R^d = \text{Im}(P_r) \oplus \text{Ker}(P_r),
$$
where the subspaces $\text{Im}(P_r)$ and $\text{Ker}(P_r)$ are not necessarily orthogonal unless $P_r$ is orthogonal. The fact that the data are only informative in $\text{Im}(P_r)$ means there exists an approximation to  the posterior density $\pi_\text{pos}^y (x)\propto\mathcal{L}^y( x) \pi_\text{pr}(x)$ under the form
\begin{equation}\label{eq:ApproximatePosterior}
 \widetilde\pi_\text{pos}^y (x) 
 \propto \widetilde{\mathcal{L}}^y( P_r x) \pi_\text{pr}(x),
\end{equation}
in which the likelihood function $x\mapsto \mathcal{L}^y(x)$ is replaced by a ridge function $x\mapsto \widetilde{\mathcal{L}}^y( P_r x)$. A ridge function \cite{pinkus2015ridge} is a function which is constant on a subspace, here $\text{Ker}(P_r)$. Let $x_r=P_rx$ and $x_\perp = (I_d-P_r)x$ be the components of $x$ in $\text{Im}(P_r)$ and $\text{Ker}(P_r)$, respectively. We have the parameter decomposition 
$$
 x = x_r + x_\perp .
$$
Using a slight abuse of notation, we factorize the prior density as $\pi_\text{pr}(x)=\pi_\text{pr}(x_r)\pi_\text{pr}(x_\perp|x_r)$, where 
\[
\pi_\text{pr}(x_r) = \int_{\text{Ker}(P_r)} \pi_\text{pr}(x_r+x_\perp')\d x_\perp' \quad {\rm and} \quad \pi_\text{pr}(x_\perp|x_r) = \pi_\text{pr}(x_r+x_\perp)/\pi_\text{pr}(x_r)
\]
denote the marginal prior and the conditional prior. The approximate posterior \eqref{eq:ApproximatePosterior} writes
$$
 \widetilde\pi_\text{pos}^y ( x_r + x_\perp) 
 \propto  \underbrace{\big( \widetilde{\mathcal{L}}^y( x_r ) \pi_\text{pr}(x_r) \big)}_{\widetilde\pi_\text{pos}^y(x_r)} \pi_\text{pr}(x_\perp|x_r).
$$
This factorization shows that, under the approximate posterior density, the Bayesian update is effective on the informed subspace $\text{Im}(P_r)$ (first term $\widetilde\pi_\text{pos}^y(x_r)$), while the non-informed subspace $\text{Ker}(P_r)$ is characterized by the prior (second term $\pi_\text{pr}(x_\perp|x_r)$). This property will be exploited later on to design efficient sampling strategies for exploring both the approximate posterior and the full posterior.

The challenge of dimension reduction is to construct both the low-rank projector $P_r$ and the ridge approximation $\widetilde{\mathcal{L}}^y$ such that the KL divergence of the full posterior from its approximation
$$
 \Dkl(\pi_\text{pos}^y||\widetilde\pi_\text{pos}^y) = \int \log\left(\frac{\pi_\text{pos}^y(x) }{ \widetilde\pi_\text{pos}^y(x)}\right)\pi_\text{pos}^y(x)\d x ,
$$
can be controlled. In this work, we specifically focus on constructing a projector $P_r$ which is independent on the data $y$ and which allows to bound $\Dkl(\pi_\text{pos}^y||\widetilde\pi_\text{pos}^y)$.

\section{Dimension reduction via optimal parameter-reduced likelihood}\label{sec3}

In this section, we first briefly review the optimal parameter-reduced likelihood and the data-dependent LIS proposed in \cite{zahm2018certified}, and then we will introduce the data-free LIS. 

\subsection{Optimal parameter-reduced likelihood using a given projector}\label{sec:Loptimal}

As shown in Section 2.1 of \cite{zahm2018certified}, for a given data set $y$ and a given projector $P_r$, the parameter-reduced likelihood function
\begin{equation}\label{eq:Loptimal}
 \mathcal{L}^{\ast,y}(x_r) = \int_{\text{Ker}(P_r)} \mathcal{L}^y( x_r + x_\perp ) \pi_\text{pr}(x_\perp|x_r) \, \d x_\perp,
\end{equation}
is an optimal approximation in the sense that it minimizes $\widetilde{\mathcal{L}}^y \mapsto \Dkl(\pi_\text{pos}^y||\widetilde\pi_\text{pos}^y)$. We denote by
$$
 \pi_\text{pos}^{\ast,y}(x) \propto \mathcal{L}^{\ast,y}(P_r x) \pi_\text{pr}(x) ,
$$
the resulting approximate posterior density. The marginal density $\pi_\text{pos}^{\ast,y}(x_r) = \int_{\text{Ker}(P_r)} \pi_\text{pos}^{\ast,y}(x_r+x_\perp)\d x_\perp$ can be expressed as
\begin{equation}\label{eq:MarginalPosteriorAgree}
 \pi_\text{pos}^{\ast,y}(x_r)
 \propto \mathcal{L}^{\ast,y}(x_r) \pi_\text{pr}(x_r)
 \overset{\eqref{eq:Loptimal}}{\propto} \int_{\text{Ker}(P_r)} \mathcal{L}^y(x_r + x_\perp)\pi_\text{pr} (x_r + x_\perp) \d x_\perp 
 \overset{\eqref{eq:BayesianModel}}{\propto} 
 \pi_\text{pos}^y(x_r) ,
\end{equation}
for all $x_r\in\text{Im}(P_r)$, where $\pi_\text{pos}^y(x_r) = \int_{\text{Ker}(P_r)} \pi_\text{pos}^y(x_r + x_\perp') \d x_\perp'$ is the marginal density of the full posterior.
Thus, for any projector $P_r$ and any data $y$, the approximate posterior $\pi_\text{pos}^{\ast,y}$ and the full posterior $\pi_\text{pos}^y$ have the same marginal density on $\text{Im}(P_r)$.
In summary we have 
\begin{align*}
\pi_\text{pos}^y(x) & = \pi_\text{pos}^y( x_r)\pi_\text{pos}^y(x_\perp | x_r) ,\\
\pi_\text{pos}^{\ast,y}(x) & \overset{\eqref{eq:MarginalPosteriorAgree}}{=} \pi_\text{pos}^y( x_r) \pi_\text{pr}(x_\perp | x_r) ,
\end{align*}
which shows that the optimal approximation $\pi_\text{pos}^{\ast,y}(x)$ to $\pi_\text{pos}^y(x)$ replaces the conditional posterior $\pi_\text{pos}^y(x_\perp | x_r)$ with the conditional prior $\pi_\text{pr}(x_\perp | x_r)$.

\subsection{Data-dependent dimension reduction}\label{sec:DataDependentDR}

We denote by $P_r=P_r^y$ a projector built by a data-dependent approach. Ideally, we would like to build $P_r^y$ that minimizes $\Dkl(\pi_\text{pos}^y||\pi_\text{pos}^{\ast,y})$ over the manifold of rank-$r$ projectors.
However, this non-convex minimization problem can be challenge to solve. Instead, the strategy proposed in \cite{zahm2018certified} minimizes an upper bound of the KL divergence obtained by logarithmic Sobolev inequalities, in which the following assumption on the prior density is adopted.

\begin{assumption}[Subspace logarithmic Sobolev inequality]\label{assu:SubspaceLogSob}
 There exists a symmetric positive definite matrix $\Gamma\in\R^{d\times d}$ and a scalar $\kappa>0$ such that for any projector $P_r\in\R^{d\times d}$ and for any continuously differentiable function $h:\R^d\rightarrow \R$ the inequality
 \begin{align*}
  \int_{\R^d} h(x)^2\log \left(\frac{h(x)^2}{ h_{P_r}(x)^2 } \right)\pi_\text{pr}(x)  \d x 
  \leq 2\kappa \int_{\R^d}\| (I_d-P_r)^\top\nabla h(x) \|_{\Gamma^{-1}}^2 \pi_\text{pr}(x)\d x ,
 \end{align*}
 holds, where $h_{P_r}^2$ is the conditional expectation of $h^2$ given by $ h_{P_r}(x)^2= \int_{\text{Ker}(P_r)} h( P_r x + x_\perp )^2 \pi_\text{pr}(x_\perp|P_r x) \, \d x_\perp$.
 Here the norm $\|\cdot\|_{\Gamma^{-1}}$ is defined by $\|v\|_{\Gamma^{-1}}^2=v^\top\Gamma^{-1}v$ for any $v\in\R^d$.
\end{assumption}

Theorem 1 in \cite{zahm2018certified} gives sufficient conditions on the prior density such that Assumption \ref{assu:SubspaceLogSob} holds. In particular, any Gaussian prior $\pi_\text{pr}=\mathcal{N}(m_\text{pr},\Sigma_\text{pr})$ with mean $m_\text{pr}\in\R^d$ and non-singular covariance matrix $\Sigma_\text{pr}\in\R^{d\times d}$ satisfies Assumption \ref{assu:SubspaceLogSob} with $\kappa=1$ and $\Gamma = \Sigma_\text{pr}^{-1}$. As shown in \cite[example 2]{zahm2018certified}, any Gaussian mixture also satisfies this assumption, but with a constant $\kappa$ which might not be accessible in practice.
We refer to \cite{guionnet2003lectures,ledoux2001logarithmic} for nicely written introductions to logarithmic Sobolev inequalities and examples of distributions which satisfy it.

\begin{proposition}\label{prop:BoundKL}
Suppose $\pi_\text{pr}$ satisfies Assumption \ref{assu:SubspaceLogSob} and the likelihood function $\mathcal{L}^y$ is continuously differentiable.
 Then, for any projector $P_r\in\R^{d\times d}$, the posterior approximation $\pi_\text{pos}^{\ast,y}(x)\propto \big( \mathcal{L}^{\ast,y}(x_r) \pi_\text{pr}(x_r) \big) \pi_\text{pr}(x_\perp|x_r)$ induced by the optimal parameter-reduced likelihood as in \eqref{eq:Loptimal} satisfies
 \begin{equation}\label{eq:ControlDkl}
  \Dkl( \pi_\text{pos}^y || \pi_\text{pos}^{\ast,y} ) \leq  \frac{\kappa}{2}  \trace\big(\Gamma^{-1} (I_d - P_r^\top )  H(y) (I_d-P_r) \big),
 \end{equation}
 where the matrix $H(y)\in\R^{d\times d}$ is defined by
 \begin{equation}\label{eq:defHy}
  H(y) = \int_{\R^d}  \big( \nabla \log\mathcal{L}^y(x) )( \nabla \log\mathcal{L}^y(x) \big)^\top \, \pi_\text{pos}^y(x) \d x.
 \end{equation}
\end{proposition}

\begin{proof}
 See the proof of Corollary 1 in \cite{zahm2018certified}.
\end{proof}

Proposition \ref{prop:BoundKL} gives an upper bound on $\Dkl( \pi_\text{pos}^y || \pi_\text{pos}^{\ast,y})$. The minimizer of this bound
$$
 P_r^y \in \underset{\substack{ P_r\in\R^{d\times d} \\ \text{rank-$r$ projector}}}{\text{arg\,min}} \,
 \trace\big(\Gamma^{-1} (I_d - P_r^\top )  H(y) (I_d-P_r) \big),
$$
can be obtained from the leading generalized eigenvectors of the matrix pair $(H(y),\Gamma)$, see \cite[Proposition 2.6]{zahm2018gradient}.
Let $(\lambda_i^y,v_i^y)\in \R_{\geq0}\times\R^d$ denotes the $i$-th eigenpair of $(H(y),\Gamma)$ such that
$
 H(y) v_i^y = \lambda_i^y \Gamma v_i^y,
$
with $(v_i^y)^\top\Gamma v_j^y = \delta_{i,j}$ and $\lambda_i^y \geq \lambda_j^y$ for all $i\leq j$. The image and the kernel of $P_r^y$ are respectively defined as
\begin{equation}\label{eq:PrY}
\begin{aligned}
 \text{Im}(P_r^y) &= \text{span}\{v_1^y,\hdots,v_r^y\}, \\
 \text{Ker}(P_r^y) &= \text{span}\{v_{r+1}^y,\hdots,v_d^y\}.
\end{aligned}
\end{equation}
The resulting projector $P_r^y$ yields an approximate posterior density $\pi_\text{pos}^{\ast,y}$ that satisfies
$$
 \Dkl( \pi_\text{pos}^y || \pi_\text{pos}^{\ast,y}) \leq \frac{\kappa}{2}  \sum_{i=r+1}^d \lambda_i^y.
$$
The above relation can be used to choose the rank $r=\text{rank}(P_r^y)$ to guarantee that the $\Dkl( \pi_\text{pos}^y || \pi_\text{pos}^{\ast,y})$ is bounded below some user-defined tolerance. A rapid decay in the spectrum $(\lambda_1^y,\lambda_2^y,\hdots)$ ensures that one can choose a rank $r$ that is much lower than the original dimension $d$.
Note that the projector $P_r^y$ may not be unique, unless there exists a spectral gap $\lambda_{r}^y>\lambda_{r+1}^y$ which ensures the $r$-dimensional dominant eigenspace of $(H(y),\Gamma)$ is unique.

\begin{remark}[Coordinate selection]\label{rmk:CoordinateSelection}
 The projector defined in \eqref{eq:PrY} is, in general, not aligned with the canonical coordinates. However, in some parametrizations---for example, different components of $x$ represent physical quantities of different nature---we may prefer coordinate selection than subspace identification to make the dimension reduction more interpretable.
 Denoting the $i$-th canonical basis vector of $\R^d$ by $e_i$, we let
 $
  P_r^y = \sum_{i\in\mathcal{I}} e_ie_i^\top,
 $
 be the projector of rank $r=\#\mathcal{I}$, which extracts the components of $x$ indexed by the index set $\mathcal{I}\subset\{1,\hdots,d\}$ 
 such that
 \begin{align*}
   \text{Im}(P_r^y) &= \text{span}\{x_i : i\in\mathcal{I}\}, \\
 \text{Ker}(P_r^y) &= \text{span}\{x_i : i\notin\mathcal{I}\}.
 \end{align*}
 Using such a projector, the bound \eqref{eq:ControlDkl} becomes
 \begin{equation*}
  \Dkl( \pi_\text{pos}^y || \pi_\text{pos}^{\ast,y} ) \leq  \frac{\kappa}{2} \sum_{i\notin\mathcal{I}}
  (\Gamma^{-1})_{ii} H(y)_{ii},
 \end{equation*}
 which suggests to define the index set $\mathcal{I}$ that selects the $r$ largest values of $(\Gamma^{-1})_{ii} H(y)_{ii}$.
\end{remark}

Because of the dependency on the data set $y$, the projector $P_r^y$ must be built after a data set has been observed, see Algorithm \ref{alg:DataDependentDR}. For scenarios where one wants to solve multiple inverse problems with multiple data sets, the matrix $H(y)$ and the resulting projector have to be reconstructed for each data set. This can be a computationally challenging task. In addition, $H(y)$ is defined as an expectation over the high-dimensional posterior distribution, which further raises the computational burden.

\begin{algorithm}[h]
\SetKwInput{Requires}{Requires}
\SetKwInput{Return}{Return}
\SetKwBlock{Offline}{Offline phase}{end}
\SetKwBlock{Online}{Online phase: \textnormal{given the data $y$ \textbf{do:}}}{{\Return{\normalfont Approximate posterior $\pi_\text{pos}^{\ast,y} (x) \propto \mathcal{L}^{\ast,y}( P_r^y x) \pi_\text{pr}(x)$.}}\vspace{-0.4cm}}
\Requires{$\pi_\text{pr}$ satisfying Assumption \ref{assu:SubspaceLogSob},  tolerance $\varepsilon>0$ and maximal rank $r_{\max}$}
\Online{
Compute the matrix $H(y)$ using \eqref{eq:defHy}.\\
Compute the generalized eigendecomposition $H(y) v_i^y = \lambda_i^y \Gamma v_i^y$.\\
Find the smallest $r$ such that $\frac{\kappa}{2}  \sum_{i=r+1}^d \lambda_i^y\leq \varepsilon$. If $r\geq r_{\max}$, set $r=r_{\max}$.\\
Assemble the projector $P_r^y$ using \eqref{eq:PrY}.\\
Define the conditional expectation $\mathcal{L}^{\ast,y}(x_r)$ defined in \eqref{eq:Loptimal}.
}
\caption{Data-dependent dimension reduction.}
\label{alg:DataDependentDR}
\end{algorithm}

\subsection{Data-free dimension reduction}\label{sec:DataFreeDR}

To overcome the abovementioned computational burden of recomputing the data-dependent projector for every new data set, we present a new data-free dimension reduction method. The key idea is to control the KL divergence in expectation over the marginal density of data. We introduce an $m$-dimensional random vector
$$
 Y \sim \pi_\text{data}(y),
$$
where $\pi_\text{data}$ is the marginal density of data defined in \eqref{eq:PiData}. Note that the observed data $y$ corresponds to a particular realization of $Y$. For a given projector $P_r$ independent on the data, replacing $y$ with $Y$ in \eqref{eq:ControlDkl} and taking the expectation over $Y$ yields
\begin{equation}\label{eq:ControlDklEXPECTATION}
 \E \big[ \Dkl( \pi_\text{pos}^Y || \pi_\text{pos}^{\ast,Y}  )  \big] 
 \leq \frac{\kappa}{2}  \trace\big(\Gamma^{-1} (I_d - P_r^\top ) \E[H(Y)] (I_d-P_r) \big).
\end{equation}
Here, the approximate posterior $\pi_\text{pos}^{\ast,Y}$ depends on $Y$ via the optimal likelihood $\mathcal{L}^{\ast,Y}$.
Similar to the data-dependent case, the leading generalized eigenvectors of the matrix pair $(\E[H(Y)], \Gamma)$ can be used to obtain a projector that minimizes the error bound. 
However, in this case, the matrix $\E[H(Y)]$ is the expectation of $H(y)$ over the marginal density of data, and thus it is independent of observed data.
Let $(\lambda_i,v_i)\in \R_{\geq0}\times \R^d$ denotes the $i$-th eigenpair of $(\E[H(Y)], \Gamma)$ such that $\E[H(Y)] v_i = \lambda_i \Gamma v_i$, with $v_i^\top\Gamma v_j = \delta_{i,j}$ and $\lambda_i^y \geq \lambda_j^y$ for all $i\leq j$. The data-free projector $P_r$ that minimizes the right-hand side of \eqref{eq:ControlDklEXPECTATION} is given by
\begin{equation}\label{eq:Pr}
\begin{aligned}
 \text{Im}(P_r) &= \text{span}\{v_1,\hdots,v_r\}, \\
 \text{Ker}(P_r) &= \text{span}\{v_{r+1},\hdots,v_d\}.
\end{aligned}
\end{equation}
When using this projector for defining the approximate posterior $\pi_\text{pos}^{\ast,Y}$, the expectation of the KL divergence $\Dkl( \pi_\text{pos}^Y || \pi_\text{pos}^{\ast,Y}  )$ can be controlled as 
\begin{equation}\label{eq:KLBoundExpectation}
 \E \big[ \Dkl( \pi_\text{pos}^Y || \pi_\text{pos}^{\ast,Y}  )  \big] 
 \leq \frac{\kappa}{2}  \sum_{i=r+1}^d \lambda_i.
\end{equation}

\begin{remark}[Bound in high probability]
Inequality \eqref{eq:KLBoundExpectation} gives a bound on $\Dkl( \pi_\text{pos}^Y || \pi_\text{pos}^{\ast,Y})$ in expectation. 
In order to obtain a bound in high probability, let us use the Markov inequality $\mathbb{P}\{\Dkl( \pi_\text{pos}^Y || \pi_\text{pos}^{\ast,Y}  )  \leq  \varepsilon \} \geq 1- \varepsilon^{-1}\E[\Dkl( \pi_\text{pos}^Y || \pi_\text{pos}^{\ast,Y}  )]$ for some $\varepsilon>0$. Thus, for a given $0<\eta\leq1$, the condition $\frac{\kappa}{2}  \sum_{i=r+1}^d \lambda_i \leq \frac{\varepsilon}{\eta}$ is sufficient to ensure that
$$
  \Dkl( \pi_\text{pos}^Y || \pi_\text{pos}^{\ast,Y}  )\leq \varepsilon,
$$
holds with a probability greater than $1-\eta$.
\end{remark}

\begin{remark}[Coordinate selection]\label{rmk:CoordinateSelection2}
 Similarly to Remark \ref{rmk:CoordinateSelection}, instead of defining $P_r$ as in \eqref{eq:Pr}, we can define a coordinate-aligned projector $P_r= \sum_{i\in\mathcal{I}} e_ie_i^\top$ by selecting an index set $\mathcal{I}$ corresponding to the $r$ largest values of $(\Gamma^{-1})_{ii}\E[H(Y)]_{ii}$.
\end{remark}

Now we show that the matrix $\E[H(Y)]$ admits a simple expression in terms of the Fisher information matrix associated with the likelihood function.
This leads to a computationally convenient way to construct the data-free projector. 
Recall that the likelihood $\mathcal{L}^y(x)$, seen as a function of $y$, is the pdf of the data $y$ conditionned on the parameter $x\in\R^d$.
The Fisher information matrix associated with this family of pdf is 
\begin{equation}\label{eq:FIM}
 \mathcal{I}(x) = \int_{\R^m}  \big( \nabla \log \mathcal{L}^y(x) )( \nabla \log \mathcal{L}^y(x) \big)^\top \,\mathcal{L}^y(x)\, \d y .
\end{equation}
We can write
 \begin{align}
  \E[H(Y)] 
  &= \int_{\R^m} H(y) \pi_\text{data}(y)\d y \nonumber\\
  &\overset{\eqref{eq:defHy}}{=} \int_{\R^m} \left( \int_{\R^d}  \big( \nabla \log\mathcal{L}^y(x) )( \nabla \log\mathcal{L}^y(x) \big)^\top \pi_\text{pos}^y(x)\d x \right) \pi_\text{data}(y)\d y\nonumber\\
  &\overset{\eqref{eq:BayesianModel}}{=} \int_{\R^m\times \R^d}   \big( \nabla \log\mathcal{L}^y(x) )( \nabla \log\mathcal{L}^y(x) \big)^\top \frac{\mathcal{L}^y(x)\pi_\text{pr}(x)}{\int_{\R^d}\mathcal{L}^y(x')\pi_\text{pr}(x') \d x'}\, \pi_\text{data}(y) \d x \d y\nonumber\\
  &\overset{\eqref{eq:PiData}}{=} \int_{\R^m\times \R^d}   \big( \nabla \log\mathcal{L}^y(x) )( \nabla \log\mathcal{L}^y(x) \big)^\top \mathcal{L}^y(x)\pi_\text{pr}(x)\d x \d y\nonumber\\
&\overset{\eqref{eq:FIM}}{=} \int_{\R^d} \mathcal{I}(x) \pi_\text{pr}(x)\d x, \label{eq:HasFisher}
 \end{align}
which shows that the matrix $\E[H(Y)]$ is the expectation of the Fisher information matrix over the prior. 
This expression does not involve any expectation over the posterior density, which is a major advantage compared to the expression \eqref{eq:defHy} of the data-dependent matrix $H(y)$.
The methodology presented here is summarized in Algorithm \ref{alg:DataFreeDR}.

\begin{algorithm}[h]
\SetKwInput{Requires}{Requires}
\SetKwInput{Return}{Return}
\SetKwBlock{Offline}{Offline phase}{\Return{\normalfont  Projector $P_r$ defined by \eqref{eq:Pr}}\vspace{-0.3cm}}
\SetKwBlock{Online}{Online phase: \textnormal{given the data $y$ \textbf{do:}}}{\Return{\normalfont Approximate posterior $\pi_\text{pos}^{\ast,y} (x) \propto \mathcal{L}^{\ast,y}( P_r x) \pi_\text{pr}(x)$}\vspace{-0.4cm}}
\Requires{$\pi_\text{pr}$ satisfying Assumption \ref{assu:SubspaceLogSob}, Fisher information matrix $\mathcal{I}(x)$ of $\mathcal{L}^y$, tolerance $\varepsilon>0$, and maximal rank $r_{\max}$}
\Offline{
Compute the matrix $H^I = \int_{\R^d}\mathcal{I}(x)\pi_\text{pr}(x)\d x$. \\
Compute the generalized eigendecomposition $H^I v_i = \lambda_i \Gamma v_i$.\\
Find the smallest $r$ such that $\frac{\kappa}{2}  \sum_{i=r+1}^d \lambda_i\leq \varepsilon$. If $r\geq r_{\max}$, set $r=r_{\max}$.
}
\Online{Define $\mathcal{L}^{\ast,y}$ as the conditional expectation defined in \eqref{eq:Loptimal}.}
\caption{Data-free dimension reduction}
\label{alg:DataFreeDR}
\end{algorithm}

\begin{example}[Gaussian likelihood]\label{rmk:GaussianLikelihood}
Consider the parameter-to-data map is represented by a smooth forward model $G:\R^d\rightarrow\R^m$ and corrupted by an additive Gaussian noise $\xi_\text{obs}\sim \mathcal{N}(0,\Sigma_\text{obs})$ with non-singular covariance matrix $\Sigma_\text{obs}\in\R^{m\times m}$, i.e., 
\[
y=G(x)+\xi_\text{obs},\quad {\rm where} \quad \xi_\text{obs}\sim\mathcal{N}(0,\Sigma_\text{obs}).
\] 
The likelihood function takes the form $\mathcal{L}^y(x) = Z^{-1}\exp( -\frac{1}{2}\| G(x)-y \|_{\Sigma_\text{obs}^{-1}}^2 )$, where $Z=\sqrt{(2\pi)^m \,\text{det}(\Sigma_\text{obs})}$ is a normalizing constant.
The Slepian-Bangs formula gives an explicit expression for the Fisher information matrix $\mathcal{I}(x) = \nabla G(x)^\top \Sigma_\text{obs}^{-1} \nabla G(x)$, where
 $\nabla G(x) \in\R^{m\times d}$ denotes the Jacobian of the forward model $G(x)$. 
 By relation \eqref{eq:HasFisher} we obtain
 \begin{equation}\label{eq:H_GaussianLikelihood}
  \E[H(Y)] = \int_{\R^d} \nabla G(x)^\top \Sigma_\text{obs}^{-1} \nabla G(x) \, \pi_\text{pr}(x)\d x.
 \end{equation}
 A similar matrix was considered in \cite{cui2014likelihood} in the context of data-dependent dimension reduction. The major difference with \eqref{eq:H_GaussianLikelihood} is that, in \cite{cui2014likelihood}, the expectation is taken over the posterior density rather than over the prior.
\end{example}

\section{Dimension reduction via parameter-reduced forward model}\label{sec4}

In the previous Section \ref{sec3}, the detection of the data-free informed subspace is based on an approximation of the likelihood function. In this section, we present an alternative strategy which, under Gaussian likelihood assumption, consist in approximating the forward model instead of the likelihood itself. This approach is similar to the vector-valued extension of the AS method \cite{zahm2018gradient} and still yields error bounds for the expected KL divergence.

As in Example \ref{rmk:GaussianLikelihood}, let us start with a Gaussian likelihood of the form
\begin{equation}\label{eq:GaussianLikelihood}
 \mathcal{L}^y(x) = \frac{1}{Z}\exp\left( -\frac{1}{2}\| G(x)-y \|_{\Sigma_\text{obs}^{-1}}^2 \right),
\end{equation}
where $x\mapsto G(x)$ is a continuously differentiable forward model, $\Sigma_\text{obs}\in\R^{m\times m}$ is a non-singular covariance matrix and $Z=\sqrt{(2\pi)^m \,\text{det}(\Sigma_\text{obs})}$ a normalizing constant.
Our goal is to build a low-dimensional approximation to the likelihood \eqref{eq:GaussianLikelihood} by replacing the forward model with a ridge approximation $x\mapsto\widetilde G(P_r x)$.
That is, we look for a likelihood approximation of the form
\begin{equation}\label{eq:ApproximateGaussianLikelihood}
 \widetilde{\mathcal{L}}^y(P_r x)=\frac{1}{Z}\exp\left( -\frac{1}{2}\| \widetilde G(P_r x)-y \|_{\Sigma_\text{obs}^{-1}}^2 \right),
\end{equation}
where $P_r$ is a low-rank projector and where $\widetilde G$ is some parameter-reduced function defined over $\text{Ker}(P_r)$. In general, this approximate likelihood \eqref{eq:ApproximateGaussianLikelihood} is different than the previous one $\mathcal{L}^{\ast,y}$, see \eqref{eq:Loptimal},  and therefore $\widetilde{\mathcal{L}}^y$ might not be optimal with respect to the KL divergence as discussed in Section \ref{sec:Loptimal}.
The following proposition will guide the construction of the approximate forward model.
\begin{proposition}\label{prop:KLBoundExpectation_FORWARDMODEL}
 Consider the posterior density $\pi_\text{pos}^y(x) \propto \mathcal{L}^y(x) \pi_\text{pr}(x)$ with a Gaussian likelihood as in \eqref{eq:GaussianLikelihood}. For any approximate forward model $\widehat G:\R^d\rightarrow\R^m$, the resulting approximate likelihood $\widehat{\mathcal{L}}^y(x) = \frac{1}{Z}\exp( -\frac{1}{2}\| \widehat G(x)-y \|_{\Sigma_\text{obs}^{-1}}^2 )$ defines an approximate posterior density $\widehat\pi_\text{pos}^y(x) \propto \widehat{\mathcal{L}}^y( x) \pi_\text{pr}(x)$ such that
 $$
  \E\left[ \Dkl( \pi_\text{pos}^Y || \widehat\pi_\text{pos}^Y ) \right] 
  + \Dkl(\pi_\text{data}||\widehat\pi_\text{data})
  = \frac{1}{2} \int_{\R^d} \|G(x) - \widehat G(x) \|_{\Sigma_\text{obs}^{-1}}^2 \pi_\text{pr}(x)\d x .
 $$
 Here the expectation is taken over $Y\sim\pi_\text{data}$ and $\widehat\pi_\text{data}(y)=\int_{\R^d} \widehat{\mathcal{L}}^y(x)  \pi_\text{pr}(x) \, \d x$ is the approximate marginal density of data.
\end{proposition}

\begin{proof}
See \ref{proof:KLBoundExpectation_FORWARDMODEL}.
\end{proof}

Using an approximate forward model in the form of $\widehat G(x) = \widetilde G(P_rx)$, Proposition \ref{prop:KLBoundExpectation_FORWARDMODEL} ensures that the approximate posterior $\widetilde\pi_\text{pos}^y(x) \propto \widetilde{\mathcal{L}}^y(P_r x) \pi_\text{pr}(x)$ with $\widetilde{\mathcal{L}}^y(P_r x)$ as in \eqref{eq:ApproximateGaussianLikelihood} satisfies
\begin{equation}\label{eq:KLBoundExpectation_FORWARDMODEL}
  \E[ \Dkl( \pi_\text{pos}^Y || \widetilde\pi_\text{pos}^Y ) ] 
  \leq \frac{1}{2} \int_{\R^d} \|G(x) - \widetilde G(P_r x) \|_{\Sigma_\text{obs}^{-1}}^2 \pi_\text{pr}(x)\d x ,
\end{equation}
This suggests to construct a ridge approximation $\widetilde G(P_rx)$ to $G(x)$ in the $L_{\pi_\text{pr}}^2$ sense.
To accomplish this, we follow the methodology proposed in \cite{zahm2018gradient} for the approximation of multivariate function using gradient information.
First, for any projector $P_r$, the optimal function $\widetilde G^{\ast}$ that minimizes the right-hand side of \eqref{eq:KLBoundExpectation_FORWARDMODEL} is the conditional expectation 
\begin{equation}\label{eq:Goptimal}
 \widetilde G^\ast(x_r) = \int_{\text{Ker}(P_r)} G( x_r + x_\perp ) \pi_\text{pr}(x_\perp|x_r) \, \d x_\perp.
\end{equation}
Then, similarly to  Assumption \ref{assu:SubspaceLogSob}, we assume that $\pi_\text{pr}$ satisfies the following subspace Poincaré inequality.

\begin{assumption}[Subspace Poincar\'{e} inequality]\label{assu:SubspacePoincare}
 There exists a symmetric positive definite matrix $\Gamma\in\R^{d\times d}$ and a scalar $\kappa>0$ such that for any projector $P_r\in\R^{d\times d}$ and for any continuously differentiable function $h:\R^d\rightarrow \R$ the inequality
 \begin{align*}
  \int_{\R^d} \big( h(x) - h_{P_r}(x) \big)^2 \pi_\text{pr}(x) \d x 
  \leq \kappa \int_{\R^d}\| (I_d-P_r)^\top\nabla h(x) \|_{\Gamma^{-1}}^2 \pi_\text{pr}(x)\d x ,
 \end{align*}
 holds, where $h_{P_r}$ is the conditional expectation of $h$ defined by $ h_{P_r}(x)= \int_{\text{Ker}(P_r)} h( P_r x + x_\perp' ) \pi_\text{pr}(x_\perp'|P_r x)  \d x_\perp'$.
\end{assumption}

Assumption \ref{assu:SubspacePoincare} is weaker than Assumption \ref{assu:SubspaceLogSob}, in the sense that any distribution which satisfies the subspace logarithmic Sobolev inequality automatically satisfies the subspace Poincar\'{e} inequality with the same $\kappa$ and the same $\Gamma$, see for instance \cite[Corollary 2]{zahm2018certified}. We refer to the recent contributions \cite{bebendorf2003note,parente2020generalized,roustant2017poincare} for examples of probability distribution which satisfy (subspace) Poincar\'{e} inequality. As for the logarithmic-Sobolev constant, the Poincar\'e constant is hard to compute in practice, except the case of Gaussian prior.
Using similar arguments as in the proof of Proposition 2.5 in \cite{zahm2018gradient}, Assumption \ref{assu:SubspacePoincare} allows to write
\begin{equation}\label{eq:KLBoundExpectation_TER}
 \int_{\R^d} \|G(x) - \widetilde G^\ast(P_r x) \|_{\Gamma_\text{obs}^{-1}}^2 \pi_\text{pr}(x)\d x
 \leq \kappa \, \trace\big(\Gamma^{-1} (I_d - P_r^\top ) H^G (I_d-P_r) \big),
\end{equation}
holds for any projector $P_r$, where the matrix $H^G\in\R^{d\times d}$ is defined by
\begin{equation}\label{eq:H_GaussianLikelihood_BIS}
  H^G = \int_{\R^d} \nabla G(x)^\top \Sigma_\text{obs}^{-1} \nabla G(x) \, \pi_\text{pr}(x)\d x,
\end{equation}
with $\nabla G(x)$ the Jacobian matrix of $G(x)$ given by
$$
 \nabla G(x) =
 \begin{pmatrix}
  \frac{\partial G_1}{\partial x_1}(x) &\hdots& \frac{\partial G_1}{\partial x_d}(x) \\
  \vdots&\ddots&\vdots \\
  \frac{\partial G_m}{\partial x_1}(x) &\hdots& \frac{\partial G_m}{\partial x_d}(x) \\
 \end{pmatrix}.
$$
Again, the projector $P_r^G$ that minimizes the right-hand side of \eqref{eq:KLBoundExpectation_TER} can be constructed via the generalized eigenvalue problem $H^Gv_i^G = \lambda_i^G \Gamma v_i^G$:
\begin{equation}\label{eq:PrG}
\begin{aligned}
 \text{Im}(P_r^G) &= \text{span}\{v_1^G,\hdots,v_r^G\}, \\
 \text{Ker}(P_r^G) &= \text{span}\{v_{r+1}^G,\hdots,v_d^G\}.
\end{aligned}
\end{equation}
Using this projector to construct the approximate forward model $\widetilde G^\ast$ in \eqref{eq:Goptimal} and the approximate likelihood as in \eqref{eq:ApproximateGaussianLikelihood}, Proposition \ref{prop:KLBoundExpectation_FORWARDMODEL} and the inequality in \eqref{eq:KLBoundExpectation_TER} yield
\begin{equation}\label{eq:KLBoundExpectation_BIS}
 \E \big[ \Dkl( \pi_\text{pos}^Y || \widetilde\pi_\text{pos}^Y  )  \big] 
 \leq \frac{\kappa}{2}  \sum_{i=r+1}^d \lambda_i^G.
\end{equation} 
The methodology is summarized in Algorithm \ref{alg:DRofForwardModel}.

The matrix $H^G$ used in this case takes the same form as the matrix $\E[H(Y)]$ in Section \ref{sec:DataFreeDR} with the Gaussian likelihood (cf. Example \ref{rmk:GaussianLikelihood}), and hence results in the same data-free projector. However, the resulting approximate likelihood functions are not the same. Indeed in Section \ref{sec:DataFreeDR}, the optimal approximate likelihood $\mathcal{L}^{\ast,y}$ is given as the conditional expectation of the likelihood function (cf. \eqref{eq:Loptimal}), whereas here, $\widetilde{\mathcal{L}}^y$ is defined by the conditional expectation of the forward model $\widetilde{G}^\ast$ (cf. \eqref{eq:Goptimal}). Using either the parameter-reduced likelihood in \eqref{eq:Loptimal} or the parameter-reduced forward model in \eqref{eq:Goptimal} results in the same parameter truncation error bound in terms of expected KL divergence.

\begin{algorithm}[H]
\SetKwInput{Requires}{Requires}
\SetKwInput{Return}{Return}
\SetKwBlock{Offline}{Offline phase}{\Return{\normalfont  Approximate forward model $x\mapsto \widetilde G^\ast (P_r^G x)$}\vspace{-0.3cm}}
\SetKwBlock{Online}{Online phase: \textnormal{given the data $y$ \textbf{do:}}}{\Return{\normalfont Approximate posterior $\widetilde\pi_\text{pos}^y (x) \propto \widetilde{\mathcal{L}}^y( P_r x) \pi_\text{pr}(x)$}\vspace{-0.4cm}}
\Requires{$\pi_\text{pr}$ satisfying Assumption \ref{assu:SubspaceLogSob}, Jacobian $\nabla G(x)$, tolerance $\varepsilon>0$ and maximal rank $r=r_{\max}$.}
\Offline{
Compute the matrix $H^G$ defined in \eqref{eq:H_GaussianLikelihood_BIS} \\
Compute the generalized eigendecomposition $H^Gv_i^G = \lambda_i^G \Gamma v_i^G$\\
Find the smallest $r$ such that $\frac{\kappa}{2}  \sum_{i=r+1}^d \lambda_i\leq \varepsilon$. If $r\geq r_{\max}$, set $r=r_{\max}$\\
Assemble the projector $P_r^G$ defined in \eqref{eq:PrG}\\
Define $\widetilde G^\ast$ as the conditional expectation \eqref{eq:Goptimal}
}
\Online{
Assemble $\widetilde{\mathcal{L}}^y(P_r^Gx)$ as in \eqref{eq:ApproximateGaussianLikelihood}
}
\caption{Data-free dimension reduction via forward model approximation}
\label{alg:DRofForwardModel}
\end{algorithm}

\begin{remark}
Despite the similarity between the approximate likelihood functions given in \eqref{eq:Loptimal} and \eqref{eq:ApproximateGaussianLikelihood}, these two approaches offer different computational characteristics. 
Given the data-free informed subspace, the optimal parameter-reduced forward model $x_r\mapsto G^\ast (x_r)$ can be further replaced by a surrogate model $x_r\mapsto G^\text{ROM}(x_r)$ constructed in the offline phase. The surrogate model can be obtained using tensor methods \cite{billaud2014tensor,Nouy2017}, the reduced basis method \cite{patera2007reduced}, polynomial techniques \cite{le2010spectral}, etc., just to cite a few. All these approximation techniques do not scale well with the apparent parameter dimensions $d$, and thus parameter reduction can greatly improve the scalability of surrogate models.

In contrast, the conditional expectation of the likelihood function in \eqref{eq:Loptimal} cannot be replaced with offline surrogate models because of the data-dependency of the likelihood. 
\end{remark}

\section{Sampling the approximate posterior}\label{sec5}

Given a data-free informed subspace, the approximate posterior density has the factorized form
\begin{equation}\label{eq:factorizationOfApproxPost}
 \widetilde\pi_\text{pos}^y (x) \propto  \widetilde\pi_\text{post}^y(x_r)  \pi_\text{pr}(x_\perp|x_r),
\end{equation}
with either $\widetilde\pi_\text{post}^y(x_r)= \pi_\text{post}^y(x_r)$ in the optimal parameter-reduced likelihood approach of Section \ref{sec3}, or with $\widetilde\pi_\text{post}^y(x_r)= \widetilde{\mathcal{L}}^y( x_r ) \pi_\text{pr}^y(x_r)$, 
$\widetilde{\mathcal{L}}^y( x_r ) \propto \exp(-\frac{1}{2}\|\widetilde G^\ast(x_r) -y\|_{\Sigma_\text{obs}^{-1}}^2)$ in the optimal parameter-reduced forward model approach of Section \ref{sec4}.
The factorization \eqref{eq:factorizationOfApproxPost} naturally suggests a dimension robust way to sampling the approximate posterior. 
The sampling method consists in first drawing samples $x_r^{(1)},x_r^{(2)},\ldots,x_r^{(K)}$ from the low-dimensional density $\widetilde\pi_\text{pos}^y ( x_r)$ using either MCMC or SMC method.
Then, for each sample $x_r^{(j)}$, we simulate a conditional prior sample $x_\perp^{(j)}$ from $\pi_\text{pr}(x_\perp|x_r^{(j)})$. In the end, $x^{(j)} = x_r^{(j)}+x_\perp^{(j)}$ are samples from the approximate posterior $\widetilde\pi_\text{pos}^y (x)$.

We emphasis here that the key is to be able to sample from the conditional prior $\pi_\text{pr}(x_\perp|x_r)$. This task is rather easy for Gaussian priors. We show in Section \ref{sec_besov} how to sample from $\pi_\text{pr}(x_\perp|x_r)$ for non-Gaussian priors with a particular structure that can be exploited.

\begin{remark}
 If the end goal is to compute expectation of some function $h$ over of the approximate posterior, the factorization \eqref{eq:factorizationOfApproxPost} leads to 
 \begin{align*}
  \int_{\R^d} h(x)\widetilde\pi_\text{pos}(x)\d x 
  &= \int_{\text{Im}(P_r)} \left(\int_{\text{Ker}(P_r)} h(x_r+x_\perp) \pi_\text{pr}(x_\perp|x_r)\d x_\perp \right) \widetilde\pi_\text{pos}(x_r)\d x_r \\
  &\approx \frac{1}{K}\sum_{j=1}^K \int_{\text{Ker}(P_r)} h(x_r^{(j)}+x_\perp) \pi_\text{pr}(x_\perp|x_r^{(j)})\d x_\perp,
 \end{align*}
 where $x_r^{(1)},\ldots,x_r^{(K)}$ are samples from the approximate marginal posterior $\widetilde\pi_\text{pos}^y(x_r)$. This way, if the expectation over the conditional prior $\pi_\text{pr}(x_\perp|x_r^{(j)})$ can be carried out analytically, one can can simply avoid using conditional prior samples. 
 Alternatively, the $K$ conditional expectations $\int h(x_r^{(j)}+x_\perp) \pi_\text{pr}(x_\perp|x_r^{(j)})\d x_\perp$ can also be approximated via other accurate quadrature rule for $\pi_\text{pr}(x_\perp|x_r^{(j)})$.
 Either way, we assume that integration with respect to the conditional prior is tractable.
\end{remark}

In Algorithm \ref{alg:MCMC_approx} we provide the details of an MCMC-based sampling procedure in which the approximate likelihood (defined by either optimal parameter-reduced likelihood or optimal parameter-reduced forward model) can be obtained as sample averages over the conditional prior $\pi_\text{pr}(x_\perp|x_r)$.
To make these approximations generally applicable, we replace the conditional prior with the marginal prior $\pi_\text{pr}(x_\perp)$ in computing those conditional expectations in the Equations \eqref{eq:Loptimal_sample} and \eqref{eq:Goptimal_sample} in Algorithm \ref{alg:MCMC_approx}.
Note that the typical class of inverse problems equipped with a Gaussian prior $\pi_\text{pr}=\mathcal{N}(m_\text{pr},\Sigma_\text{pr})$ is a special case. Since the projector $P_r$ is orthogonal with respect to $\Sigma_\text{pr}^{-1}$, the marginal prior $\pi_\text{pr}(x_\perp)$ coincides with the conditional prior $\pi_\text{pr}(x_\perp|x_r)$. 

\begin{algorithm}[ht!]
    \SetKwInput{Requires}{Requires}
    \SetKwInput{Return}{Return}
    \Requires{A projector $P_r$, a sample size $N$ for approximating the likelihood, a total posterior sample size $K$, and a proposal density $q(\cdot|x_r)$  on $\text{Im}(P_r)$.}

    \SetKwBlock{CD}{Sample-averaged likelihood approximation}{\Return{\normalfont a sample-averaged approximate likelihood function $\widetilde{\mathcal{L}}_N^y( x_r )$}\vspace{-0.3cm}}
    \CD{
    Draw $N$ i.i.d. samples $x_\perp^{(1)},\hdots,x_\perp^{(N)}$ from the marginal $\pi_\text{pr}(x_\perp)$\\
    \If{optimal parameter-reduced likelihood is used}{
        \begin{equation}\label{eq:Loptimal_sample}
        \widetilde{\mathcal{L}}_N^y( x_r ) = \frac1N \sum_{i = 1}^{N} \mathcal{L}^y( x_r + x_\perp^{(i)} ) \vspace{-0.5cm}
        \end{equation}
        for any $x_r\in\text{Im}(P_r)$
    }
    \If{optimal parameter-reduced forward model is used}{
    \begin{equation}\label{eq:Goptimal_sample}
     \widetilde{\mathcal{L}}_N^y( x_r ) \propto \exp\Big(-\frac{1}{2}\|\widetilde{G}_N(x_r) -y\|_{\Sigma_\text{obs}^{-1}}^2\Big), \widetilde{G}_N(x_r) = \frac1N \sum_{i = 1}^{N} G( x_r + x_\perp^{(i)} )
    \end{equation}
    for any $x_r\in\text{Im}(P_r)$
    }}

    \SetKwBlock{MCMC}{Subspace MCMC sampling}{\Return{\normalfont a Markov chain $X_r^{(1)},X_r^{(2)},\ldots,X_r^{(K)}$ with invariant density $\widetilde\pi_\text{pos}^y ( x_r)$  \vspace{-0.3cm}}}
    \MCMC{
    \For{$j = 1,2,\ldots, K$}{
        Given the Markov chain state $X_r^{(j-1)} = x_r$, propose a candidate $x_r^\dagger \sim q(\cdot|x_r)$ \\
        Evaluate the approximate likelihood function $\widetilde{\mathcal{L}}_N^y( x_r^\dagger )$\\
        Compute the acceptance probability \vspace{-0.2cm}
        \begin{equation}\label{eq:alpha_algo}
         \alpha(x_r^\dagger|x_r) = \min\bigg\{ 1 \,,\, \frac{\widetilde{\mathcal{L}}_N^y( x_r^\dagger ) \pi_\text{pr}(x_r^\dagger)\,q(x_r | x_r^\dagger)}{\widetilde{\mathcal{L}}_N^y( x_r )\pi_\text{pr}(x_r)\, q(x_r^\dagger | x_r)} \bigg\}.\vspace{-0.2cm}
        \end{equation}

        With probability $\alpha(x_r^\dagger|x_r)$, {\bf accept} $x_r^\dagger$ by setting $X_r^{(j)} = x_r^\dagger$, otherwise {\bf reject} $x_r^\dagger$ by setting $X_r^{(j)} = x_r$.
    
    }}

    \SetKwBlock{null}{Approximate posterior sampling}{\Return{\normalfont approximate marginal posterior samples $x^{(1)},x^{(2)},\ldots,x^{(K)}$}\vspace{-0.3cm}}
    \null{
    \For{$j = 1,2,\ldots, K$}{
        Given the state $X_r^{(j)} = x_r^{(j)}$, draw a conditional prior sample $x_\perp^{(j)} \sim \pi_\text{pr}(\cdot|x_r^{(j)})$ \\
        Compute the $i$-th approximate posterior sample $x^{(j)} = x_r^{(j)} + x_\perp^{(j)} $
    }}
    \caption{MCMC-based approach for sampling the approximate posterior.}
    \label{alg:MCMC_approx}
\end{algorithm}

A remaining question is how to choose the sample size $N$ for computing the conditional expectations in \eqref{eq:Loptimal_sample} and \eqref{eq:Goptimal_sample}. The following heuristic is developed based on the optimal parameter-reduced forward model. 
Consider the exact parameter-reduced forward model $\widetilde G^\ast(P_r x) = \widetilde G^\ast (x_r)$ and its sample-averaged approximation $\widetilde G_N(P_r x) =  \widetilde G_N(x_r)$. The sample-averaged approximation defines an approximate posterior density
\[
\widehat\pi_\text{pos}^y(x) \propto \exp\Big(-\frac{1}{2}\|\widetilde G_N(P_r x) -y\|_{\Sigma_\text{obs}^{-1}}^2\Big) \pi_\text{pr}(x),
\] 
that satisfies 
\begin{align}
 \E[ \Dkl( \pi_\text{pos}^Y || \widehat\pi_\text{pos}^Y ) ] 
 &\overset{\eqref{eq:KLBoundExpectation_FORWARDMODEL}}{\leq}
 \E\left[ \frac{1}{2} \int_{\R^d} \|G(x) - \widetilde G_N(P_r x) \|_{\Sigma_\text{obs}^{-1}}^2 \pi_\text{pr}(x)\d x \right] \nonumber\\
 & = \frac{1}{2} \, \E\left[\int_{\R^d} \| G(x) - \widetilde G^\ast(P_r x) \|_{\Sigma_\text{obs}^{-1}}^2 +  \|\widetilde G^\ast(P_r x) - \widetilde G_N(P_r x) \|_{\Sigma_\text{obs}^{-1}}^2 \pi_\text{pr}(x)\d x \right] \nonumber\\
 &= \left(1+\frac{1}{N}\right) \frac{1}{2}\int_{\R^d} \| G(x) - \widetilde G^\ast(P_r x) \|_{\Sigma_\text{obs}^{-1}}^2 \pi_\text{pr}(x)\d x . \label{eq:controlPiHat}
\end{align}
Here, the expectation is taken jointly over the data $Y$ and the sample $\{x_\perp^{(i)}\}_{i=1}^N$.
The above inequality directly follows from Proposition \ref{prop:KLBoundExpectation_FORWARDMODEL} and the fact that $\widetilde G^\ast(P_r x)$ is the conditional expectation of $G(x)$ over $\text{Ker}(P_r)$. 
We refer to Theorem 3.2 in \cite{constantine2014active} for more details on this derivation. Inequality \eqref{eq:controlPiHat} implies that the random approximate posterior $\widehat\pi_\text{pos}^y(x)$ can be used in place of $\widetilde\pi_\text{pos}^y(x)$, as the bounds on the expected Kullback-Leibler divergence in \eqref{eq:KLBoundExpectation_FORWARDMODEL} and \eqref{eq:controlPiHat} are comparable. In addition, this suggests that the sample size $N$ in \eqref{eq:Goptimal_sample} does not have to be large. Even with $N=1$, \eqref{eq:KLBoundExpectation_FORWARDMODEL} and \eqref{eq:controlPiHat} differs only by a factor of 2.
For the optimal parameter-reduced likelihood function, it is not obvious how to obtain a similar bound for the sampled-averaged conditional expectation in \eqref{eq:Loptimal_sample}, see for instance the result \cite[Proposition 5]{zahm2018certified}. 
In this case, we adopt the identity \eqref{eq:controlPiHat} as a heuristic.

\section{Sampling from the exact posterior}\label{sec6}

\label{sec:DRofForwardModel}

In this section, we present new strategies for sampling  the exact posterior by adding minor modifications to Algorithm \ref{alg:MCMC_approx}.

\subsection{Pseudo-marginal for the optimal parameter-reduced likelihood}\label{sec:PreudoMarginal}

For the optimal parameter-reduced likelihood approach, Algorithm \ref{alg:MCMC_approx} replaces the optimal likelihood $\mathcal{L}^{*,y}(x_r)$ with the sample-average $\widetilde{\mathcal{L}}^{y}_N(x_r)$ defined by \eqref{eq:Loptimal_sample} using frozen (fixed) samples $\{x_\perp^{(i)}\}_{i=1}^N$. This way, Algorithm \ref{alg:MCMC_approx} produces samples from an estimation to the posterior approximation $\pi_\text{pos}^{\ast,y}(x)=\pi_\text{pos}(x_r)\pi_\text{pr}(x_\perp|x_r)$. 
In this section, we first show that replacing the frozen samples with freshly drawing samples $\{x_\perp^{(i)}\}_{i=1}^N$ at each MCMC iteration yields a pseudo-marginal MCMC \cite{andrieu2009pseudo} which samples exactly from $\pi_\text{pos}^{\ast,y}(x)$.
In addition, we also show that an appropriate recycling of the data generated by this modified algorithm allows obtaining samples from the exact posterior $\pi_\text{pos}^y(x)$ itself.

We propose to modify Algorithm \ref{alg:MCMC_approx} by replacing the acceptance rate $\alpha_N(x_r^\dagger| x_r)$ in \eqref{eq:alpha_algo} with
\begin{equation}\label{eq:alphaHat}
 \widehat\alpha_N(x_r^\dagger| x_r) 
 = \min\left\{ 1 \,,\, 
 \frac{\pi_\text{pr}(x_r^\dagger)\left(\frac{1}{N}\sum_{i=1}^N \mathcal{L}^y(x_r^\dagger + x_\perp^{\dagger(i)})\right)q(x_r | x_r^\dagger)}{\pi_\text{pr}(x_r)\left(\frac{1}{N}\sum_{i=1}^N \mathcal{L}^y(x_r + x_\perp^{(i)})\right) q(x_r^\dagger | x_r)} 
 \right\}.
\end{equation}
Here, $\{x_\perp^{(i)}\}_{i=1}^N$ are i.i.d. samples from $\pi_\text{pr}(x_\perp|x_r)$ conditioned on the current state of the chain $x_r$ and $\{x_\perp^{\dagger(i)}\}_{i=1}^N$ are i.i.d. samples from $\pi_\text{pr}(x_\perp|x_r^\dagger)$ conditioned on the proposed candidate $x_r^\dagger$. Compared to the previous acceptance rate \eqref{eq:alpha_algo} where $\{x_\perp^{(i)}\}_{i=1}^N = \{x_\perp^{\dagger(i)}\}_{i=1}^N$ where frozen, the new acceptance rate \eqref{eq:alphaHat} requires to redraw fresh samples at each proposal candidate $x_r^\dagger$. This is summarized in Algorithm \ref{alg:PseudoMarginalMCMC}.

\begin{algorithm}[h]
    \SetKwInput{Requires}{Requires}
    \SetKwInput{Return}{Return}
    \SetKwInput{Accept}{Accept}
    \SetKwInput{Reject}{Reject}
    \SetKwBlock{Iter}{}{}
    \Requires{A projector $P_r$, a sample size $N$ for approximating the likelihood, a total posterior sample size $K$, and a proposal density $q(\cdot|x_r)$ on $\text{Im}(P_r)$.}
    \For{$j = 1, 2, \ldots, K$}{
    Given the previous state of the Markov chain $X_r^{(j-1)} = x_r $ and the associated set of conditional prior samples $\{X_\perp^{(j-1,i)}\}_{i=1}^N = \{x_\perp^{(i)}\}_{i=1}^N$ \vspace{1pt}\\
    Propose a candidate $x_r^\dagger \sim q(\cdot|x_r)$ \vspace{1pt}\\
    Draw $N$ independent samples $x_\perp^{\dagger(1)}, \ldots, x_\perp^{\dagger(N)} \sim \pi_\text{pr}(x_\perp | x_r^\dagger)$ \vspace{1pt}\\
    Compute the acceptance probability $\widehat\alpha(x_r^\dagger|x_r)$ as in \eqref{eq:alphaHat} \vspace{1pt}\\
    With probability $\widehat\alpha(x_r^\dagger|x_r)$, {\bf accept} $X_r^{(j)} = x_r^\dagger $ and $\{X_\perp^{(j,i)}\}_{i=1}^N =\{x_\perp^{(\dagger, i)}\}_{i=1}^N$. Otherwise {\bf reject} and set $X_r^{(j)} = x_r $ and $\{X_\perp^{(j,i)}\}_{i=1}^N =\{x_\perp^{(i)}\}_{i=1}^N$.
    \vspace{1pt}
    }
    \Return{\normalfont the Markov chain $\{(X_r^{(j)} , \{X_\perp^{(j,i)}\}_{i=1}^N )\}_{j=1}^K$}
    \caption{Pseudo-marginal MCMC for sampling the exact marginal posterior.}
    \label{alg:PseudoMarginalMCMC}
\end{algorithm}

In the next proposition we apply the analysis of pseudo-marginal MCMC \cite{andrieu2009pseudo} to show that $\pi_\text{pos}^y(x_r)$ is the invariant density of the Markov chain constructed by Algorithm \ref{alg:PseudoMarginalMCMC}. The key step is to interpret Algorithm \ref{alg:PseudoMarginalMCMC} as a classical Metropolis-Hastings algorithm that operates on the product space $\text{Im}(P_r) \times \text{Ker}(P_r)^N$. 

\begin{proposition}\label{prop:PseudoMarginalMCMC}
 Algorithm \ref{alg:PseudoMarginalMCMC} constructs an ergodic Markov chain $\{(X_r^{(j)} , \{X_\perp^{(j,i)}\}_{i=1}^N )\}_{j\geq1}$ on the product space $\text{Im}(P_r)\times \text{Ker}(P_r)^N$ with invariant density
 \begin{equation}\label{eq:piTar}
 \pi_\text{tar}^{y,N}(x_r,\{x_\perp^{(i)}\}_{i=1}^N) \propto \pi_\text{pr}(x_r) \left(\sum_{i=1}^N \mathcal{L}^y(x_r + x_\perp^{(i)}) \right)\prod_{j=1}^N\pi_\text{pr}(x_\perp^{(j)}|x_r).
 \end{equation}
 The marginal of this target density satisfies $\pi_\text{tar}^{y,N}(x_r)= \pi_\text{pos}^y(x_r)$ 
 so that the sequence $\{X_r^{(j)}\}_{j=1}^N$ is an ergodic Markov chain with the invariant density $\pi_\text{pos}^y(x_r)$. 
\end{proposition}
\begin{proof}
 See \ref{proof:prop:PseudoMarginalMCMC}.
\end{proof}

\begin{remark}[Choosing $N$ in Algorithm \ref{alg:PseudoMarginalMCMC}] The statistical performance of pseudo-marginal methods depends on the variance of the sample-averaged estimate $\frac{1}{N}\sum_{i=1}^N \mathcal{L}^y(x_r + x_\perp^{(i)})$. This variance being inversely proportional to the sample size $N$, a larger $N$ may result in better statistical efficiency of the MCMC chain.
However, the computational cost per MCMC iteration increases linearly with $N$, while the improvement of the statistical efficiency will not follow the same rate. We refer the readers to \cite{andrieu2016establishing,doucet2015efficient} for a detailed discussion on this topic and only provide an interpretation as follows. With $N\rightarrow\infty$, the Markov chain constructed by the pseudo-marginal MCMC converges to that of an idealized standard MCMC, which has the acceptance probability defined by the same proposal density and the exact evaluation of $\mathcal{L}^{*,y}(x_r)$. This way, even with a very large $N$, the statistical efficiency of the pseudo-marginal MCMC cannot be improved further beyond that of the idealized standard MCMC.
As suggested by \cite{doucet2015efficient}, the standard deviation of the logarithm of the parameter-reduced likelihood estimate, $\text{var}[\log \widetilde{\mathcal{L}}_N^y]^{\frac12}$, can be used to monitor the quality of the sample-averaged estimator. 
\end{remark}

It is remarkable to observe that, for $N=1$, the target density \eqref{eq:piTar} becomes the true posterior $\pi_\text{tar}^{y,1}(x_r,x_\perp) = \pi_\text{pos}^y(x_r+x_\perp)$. This means that Algorithm \ref{alg:PseudoMarginalMCMC} actually produces samples $x=x_r+x_\perp$ from $\pi_\text{pos}^y(x)$. 
For $N>1$, we propose to recycle the Markov chain $\{X_\perp^{(1,i)}\}_{i=1}^{N},\hdots,\{X_\perp^{(K,i)}\}_{i=1}^{N}$ produced by Algorithm \ref{alg:PseudoMarginalMCMC} in order to generate samples from the exact posterior $\pi_\text{pos}^y(x)$.
This procedure is summarized in Algorithm \ref{alg:MCMC_for_exact_inference} and a justification is provided in the following proposition. 

\begin{proposition}\label{prop:MCMC_for_exact_inference}
 Let $\{(X_r^{(j)} , \{X_\perp^{(j,i)}\}_{i=1}^N )\}_{j\geq1}$ be a Markov chain generated by Algorithm \ref{alg:PseudoMarginalMCMC}.
 For any $j\geq1$ we randomly select $X_\perp^{(j)} \in \{X_\perp^{(j,i)}\}_{i=1}^N$ according to the probability
 \begin{equation}\label{eq:x_perpStar}
  \mathbb{P}\left(X_\perp^{(j)} = X_\perp^{(j,k)} \Big| X_r^{(j)} ,\{X_\perp^{(j,i)}\}_{i=1}^N  \right) = \frac{\mathcal{L}^y(X_r^{(j)}+X_\perp^{(j,k)})}{\sum_{i=1}^N \mathcal{L}^y(X_r^{(j)}+X_\perp^{(j,i)})},
  \quad 1\leq k\leq N,
 \end{equation}
 and we let $X^{(j)} = X_r^{(j)} + X_\perp^{(j)}$.
 Then $\{X^{(j)}\}_{j\geq1}$ is a Markov chain with the exact posterior $\pi_\text{pos}^y(x)$ as invariant density.
\end{proposition}

\begin{proof}
See \ref{proof:prop:MCMC_for_exact_inference}
\end{proof}

\begin{algorithm}[h]
    \SetKwInput{Requires}{Requires}
    \SetKwInput{Return}{Return}
    \SetKwInput{Accept}{Accept}
    \SetKwInput{Reject}{Reject}
    \Requires{MCMC chain generated $\{(X_r^{(j)} , \{X_\perp^{(j,i)}\}_{i=1}^N )\}_{j=1}^K$ by Algorithm \ref{alg:PseudoMarginalMCMC}}
    \For{$j = 1, 2, \ldots, K$}{
    Subsample $X_\perp^{(j)} \in \{X_\perp^{(j,i)}\}_{i=1}^N$ according to the probability \eqref{eq:x_perpStar} \vspace{1pt}\\
    Assemble $X^{(j)}=X_r^{(j)}+X_\perp^{(j)}$
    }
    \Return{the Markov chain $\{X^{(j)}\}_{j=1}^K$ with invariant density $\pi_\text{pos}^y(x)$}
    \caption{Recycling the Markov chain generated by Algorithm \ref{alg:PseudoMarginalMCMC} to generate exact posterior samples}
    \label{alg:MCMC_for_exact_inference}
\end{algorithm}

\subsection{Delayed acceptance for the optimal parameter-reduced forward model}

For the optimal parameter-reduced forward model, the marginal density of the resulting approximate posterior does not coincide with that of the exact posterior in general. However, we can still modify the  approximate inference algorithm \ref{alg:MCMC_approx} using the delayed acceptance technique \cite{christen2005markov,liu2001monte,liu1998sequential} to explore the exact posterior. 
The delayed acceptance modifies Algorithm \ref{alg:MCMC_approx} by adding a second stage acceptance rejection within each MCMC iteration. 
Here we consider the sample-averaged likelihood $\widetilde{\mathcal{L}}_N^y( x_r )$ defined by either \eqref{eq:Loptimal_sample} (the optimal parameter-reduced likelihood) or \eqref{eq:Goptimal_sample} (the optimal parameter-reduced forward model), where the marginal prior sample set $\{x_\perp^{(i)}\}_{i = i}^{N}$ is prescribed. 
The following Proposition and Algorithm \ref{alg:DA_MCMC} detail this modification.

\begin{proposition}\label{prop:DAMCMC}
Suppose we have a proposal distribution $q(\cdot | x_r )$ defined in the parameter reduced subspace ${\rm Im}(P_r)$. We consider the following two stage Metropolis-Hastings method. In the first stage, we draw a proposal candidate $x_r^\dagger \sim q(\cdot | x_r )$.
Then, with the probability
\begin{equation}\label{eq:DA_acc1}
\alpha(x_r^\dagger|x_r) = \min\bigg\{ 1 \,,\, \frac{\widetilde{\mathcal{L}}_N^y( x_r^\dagger ) \pi_\text{pr}(x_r^\dagger)\,q(x_r | x_r^\dagger)}{\widetilde{\mathcal{L}}_N^y( x_r )\pi_\text{pr}(x_r)\, q(x_r^\dagger | x_r)} \bigg\},\vspace{-0.2cm}
\end{equation}
we move the proposal candidate $x_r^\dagger$ to the next stage. In the second stage, we draw a proposal candidate $\pi_{\rm pr}(x_\perp^\dagger | x_r^\dagger)$ in the complement subspace ${\rm Ker}(P_r)$ and then accept the pair of proposal candidates $(x_r^\dagger, x_\perp^\dagger)$ with the probability
\begin{equation}\label{eq:DA_acc2}
\beta(x_r^\dagger, x_\perp^\dagger|x_r, x_\perp) = \min\left[ 1, \frac{\mathcal{L}^y(x_r^\dagger+ x_\perp^\dagger) \,\widetilde{\mathcal{L}}_N^y(x_r)} {\mathcal{L}^y( x_r + x_\perp )\,\widetilde{\mathcal{L}}_N^y(x_r^\dagger)}\right].
\end{equation}
Then, the above procedure constructs an ergodic Markov chain with the full posterior $\pi_\text{pos}^y(x)$ as the invariant density. 
\end{proposition}

\begin{proof}
This result can be derived from the standard delayed acceptance \cite{christen2005markov}. For completeness, we provide the proof in \ref{proof:prop:DAMCMC}.
\end{proof}

\begin{algorithm}[h]
\SetKwInput{Accept}{Accept}
\SetKwInput{Reject}{Reject}
\SetKwInput{Requires}{Requires}
\SetKwInput{Return}{Return}
\Requires{A projector $P_r$, a sample-averaged likelihood approximation defined in Algorithm \ref{alg:MCMC_approx}, a total sample size $K$, and a proposal density $q(\cdot|x_r)$  on $\text{Im}(P_r)$.}
\For{$j = 1, 2, \ldots, K$}{
Given the Markov chain state $X^{(j-1)} = x_r + x_\perp$, propose a candidate $x_r^\dagger \sim q(\cdot|x_r)$ \vspace{1pt}\\
Compute the parameter-reduced likelihood $\widetilde{\mathcal{L}}_N^y(x_r^\dagger )$ using either using \eqref{eq:Loptimal_sample} or \eqref{eq:Goptimal_sample} \vspace{1pt}\\
With probability $\alpha(x_r^\dagger|x_r)$ in \eqref{eq:DA_acc1} {\bf move} $x_r^\dagger$ to the next stage as follows \vspace{1pt}\\
  \qquad Propose a candidate $x_\perp^\dagger \sim \pi_{\rm pr}(\cdot|x_r^\dagger)$ \vspace{1pt} \\
  \qquad Compute the full likelihood $\mathcal{L}^{y}(x_r^\dagger + x_\perp^\dagger)$ \vspace{1pt} \\
  \qquad With probability $\beta(x_r^\dagger, x^\dagger_\perp|x_r, x_\perp)$ in \eqref{eq:DA_acc2} {\bf accept} $(x_r^\dagger, x^\dagger)$, otherwise {\bf reject} $(x_r^\dagger, x^\dagger)$\hspace{-6pt} \vspace{1pt}\\ 
Otherwise {\bf reject} $x_r^\dagger$
}
\Accept{set $X^{j} = x_r^\dagger + x_\perp^\dagger$ \vspace{1pt}}
\Reject{set $X^{j} = X^{(j-1)}$ \vspace{1pt}}
\Return{\normalfont a Markov chain $X^{(1)},X^{(2)},\ldots,X^{(K)}$ with the invariant density $\pi_\text{pos}^y (x)$ }
\caption{Delayed acceptance MCMC for sampling the exact posterior.}
\label{alg:DA_MCMC}
\end{algorithm}

\begin{remark}
It worth to note that the delayed acceptance also opens the door to further accelerate the exact inference using surrogate models instead of the original forward model. 
The approximate likelihood $\widetilde{\mathcal{L}}_N^y( x_r)$ is deterministic and dimension reduced, which makes it possible to further approximate $\widetilde{\mathcal{L}}_N^y( x_r)$ using computationally fast surrogate models. 
In this case, the same delayed acceptance MCMC (Algorithm \ref{alg:DA_MCMC}) can still produce ergodic Markov chains that converge to the full posterior $\pi^{y}_{\rm pos}(x)$.
In contrast, the pseudo-marginal method requires an unbiased Monte Carlo estimate of the exact marginal posterior $\pi^{y}_{\rm pos}(x_r)$ at every iteration, which is not straightforward to accelerate using surrogate models.
\end{remark}

\section{Non-Gaussian priors}\label{sec_besov}

The dimension reduction techniques presented in Sections \ref{sec3} and \ref{sec4} require one to evaluate the marginal prior density $\pi_{\rm pr}(x_r)=\int_{\text{Ker}(P_r)} \pi_\text{pr}(x_r+x_\perp)\d x_\perp$ and draw samples from the conditional prior $\pi_{\rm pr}(x_\perp|x_r)=\pi_\text{pr}(x_r+x_\perp)/\pi_{\rm pr}(x_r)$. 
While these tasks are readily doable for Gaussian distributions, it might not be the case in general. 
In this section, we use Besov priors as an example to present strategies that can extend the proposed dimension reduction methods to some non-Gaussian priors.

\subsection{Besov priors}\label{sec_besov_2}

 Besov measure \cite{dashti2012besov,kolehmainen2012sparsity,saksman2009discretization} naturally appears in image reconstruction problems in which the detection of edges and interfaces is important.
 Following \cite{kolehmainen2012sparsity,saksman2009discretization}, we construct Besov priors using wavelet functions and consider functions on the one-dimensional torus $\mathbb{T} = (0,1]$. 
 Starting with a suitable compactly supported mother wavelet function $\psi_\ast \in \mathcal{L}_2(\mathbb{T})$, 
 we can define an orthogonal basis
 \[
 \psi_{j,k}(s) = 2^{\frac{j}2}\psi_\ast(2^j s - k), \quad j, k \in \mathbb{N}_{\geq 0}, \;\; k\in [0, 2^j -1].
 \]
 This way, given a smoothness parameter $r>0$ and integrability parameters $1\leq p,q\leq \infty$, a function $f:s\mapsto f(s)$ in the Besov space $\mathcal{B}^r_{pq}(\mathbb{T})$ can be written as
 \begin{equation}\label{eq:besov_f}
 f(s) = c_0 + \sum_{j = 0}^{\infty} \sum_{k = 0}^{2^j-1} 2^{-j(r + \frac12 - \frac{1}{p})} b_{j,k} \psi_{j,k}(s),
 \end{equation}
 and satisfies 
 \[
 \|f\|_{\mathcal{B}^r_{pq}} :=  \bigg( |c_0|^q + \sum_{j = 0}^{\infty} \Big(\sum_{k = 0}^{2^j-1} | b_{j,k}|^p\Big)^{\frac{q}{p}} \bigg)^{\frac1q} < \infty.
 \]
 In a Bayesian setting, we can set $p = q$ and define the Besov-$\mathcal{B}^r_{pp}$ prior with the pdf\footnote{This pdf is used for demonstrating the intuition rather than a rigorous characterization, as it is defined with respect to the (non-existent) infinite-dimensional Lebesgue measure. However, the finite-dimensional discretization of the Besov measure, which is used in numerical simulations, has a pdf in this form with respect to Lebesgue measure.}
 \begin{equation}\label{eq:besov_p}
 \pi_{\rm pr}(f) \propto \exp\Big( - \gamma \|f\|_{\mathcal{B}^r_{pp}}^p\Big),
 \end{equation}
 where $\gamma>0$ is a scale parameter. One can easily generalize the above definition of Besov priors to multidimensional settings by taking tensor products of the one-dimensional basis and associated coefficients.

 We can discretize the Besov prior by truncating the infinite sum in \eqref{eq:besov_f} to the first $D$ terms. This way, collecting all the coefficients into a parameter vector $x = (c_0, b_{0,0}, b_{0,1}, \ldots, b_{1,0},b_{1,1}, \ldots) \in \R^d$, where $d = 2^{D+1}$, the discretized Besov-$\mathcal{B}^r_{pp}$ prior can be equivalently expressed as a product-form distribution over the parameter $x$ with the pdf
 \begin{equation}\label{eq:PriorWithIndepCompo}
  \pi_{\rm pr} (x)= \prod_{i=1}^d \pi_{\rm pr}^{(i)}(x_i)\quad {\rm with} \quad \pi_{\rm pr}^{(i)}(x_i) \propto \exp\left( - \gamma |x_i|^p\right).
 \end{equation}

 \subsection{Dimension reduction via coordinate selection}\label{sec_coor_select}
 In general, we do not have closed form expressions for both the marginal $\pi_{\rm pr}(x_r)$ and the conditional $\pi_{\rm pr}(x_\perp|x_r)$, unless the projector $P_r$ is aligned with the canonical basis. 
 This leads to the construction of reduced subspace by selecting a subset of canonical basis. 
 As discussed in Remarks \ref{rmk:CoordinateSelection} and \ref{rmk:CoordinateSelection2}, this task can be achieved by identifying an index set $\mathcal{I}\subset\{1,\hdots,d\}$ with cardinality $r$ such that $\mathcal{I}$ contains the indices of the $r$ largest values of $i\mapsto (\Gamma^{-1})_{ii}\E[H(Y)]_{ii}$ in the data-dependent case or those of $i\mapsto (\Gamma^{-1})_{ii}\E[H(Y)]_{ii}$ in the data-free case.
 This leads to the projector 
 $
  P_r = \sum_{i\in\mathcal{I}} e_ie_i^\top,
 $
 where $\{e_1,\hdots,e_d\}$ is the canonical basis of $\R^d$.
 Thus, the product-form of \eqref{eq:PriorWithIndepCompo} yields the marginal prior and the conditional prior
 \begin{align*}
  \pi_{\rm pr} (x_r) = \prod_{i\in\mathcal{I}} \pi_{\rm pr}^{(i)}(x_i) \quad {\rm and} \quad
  \pi_{\rm pr} (x_\perp|x_r) = \prod_{i\notin\mathcal{I}} \pi_{\rm pr}^{(i)}(x_i) ,
 \end{align*}
 respectively. In this formulation, evaluating the marginal prior density and drawing samples from the conditional prior become  straightforward tasks.
\begin{remark}
 For $1\leq q<2$, the tails of $\pi_{\rm pr}^{(i)}(x_i)$ defined in \eqref{eq:PriorWithIndepCompo} are heavier than Gaussian tails, and hence Assumptions \ref{assu:SubspaceLogSob} and \ref{assu:SubspacePoincare} may not be satisfied. 
 Nonetheless, one can still numerically apply the proposed dimension reduction methods without having the error bounds in \eqref{eq:KLBoundExpectation} and \eqref{eq:KLBoundExpectation_BIS}. In this case, we set $\Gamma$ to be the identity matrix in accordance with the fact that the prior components are independent and identically distributed.
\end{remark}

\begin{remark}[Other sparsity-inducing prior]
 There exist other shrinkage priors similar to Besov priors, in which the random function is expressed as a weighted linear combination of basis functions and the associated random weights follow other type of heavy tail distributions. For example, the horseshoe prior and the Student's $t$ prior. See \cite{carvalho2009handling} for further discussions and references therein. The coordinate selection technique introduced here may also be applicable to those shrinkage priors. 

\end{remark}

\subsection{Dimension reduction via prior normalization}\label{sec_normalization}

Alternatively, we consider the case where the prior can be defined as the pushforward of the standard Gaussian measure with pdf $\mu(x)\propto\exp(-\frac12\|x\|_2^2)$ under a $C^1$-diffeomorphism $T:\R^d\rightarrow\R^d$, which takes the form
\begin{equation}\label{eq:PriorAsPushforwardOfGaussian}
\pi_{\rm pr}(x) = T_\sharp\mu(x).
\end{equation}
In other words, $\pi_{\rm pr}(x)$ is the pdf of the random vector $X=T(Z)$ where $Z\sim\mu(z)$.
For the Besov-$\mathcal{B}^r_{pp}$ prior defined in \eqref{eq:PriorWithIndepCompo}, the diffeomorphism $T$ has a diagonal form $T(z) = (T_1(z_1), \hdots, T_d(z_d))$ with $T_i(z_i)  = \Phi_i^{-1}( \Psi(z_i) )$, 
where $\Phi_i(\cdot)$ is the cumulative density function (cdf) of $\pi_{\rm pr}^{(i)}(x_i)$ defined in \eqref{eq:PriorWithIndepCompo} and $\Psi(\cdot)$ is the cdf of the standard Gaussian. We provide details of the cdf $\Phi(\cdot)$ in \ref{sec:cdf_Besov}. 

The invertibility of $T$ allows us to reparametrize the Bayesian inverse problem in terms of the variable $z=T^{-1}(x)$, which is endowed with the Gaussian prior $\mu$. With this change of variable, the likelihood function becomes $z\mapsto\mathcal{L}^y(T(z))$, and thus the matrix $H(y)$ used to reduce the dimension of $z$ should be
\begin{equation*}\label{eq:defHy_new}
 H_z(y) = \int_{\R^d}  \nabla T(z)^\top \big( \nabla \log\mathcal{L}^y(T(z)) )( \nabla \log\mathcal{L}^y(T(z)) \big)^\top \nabla T(z) \, \mu(z) \d z,
\end{equation*}
in the data-dependent case and $\E[H_z(Y)]$ in the data-dependent case.
For the optimal parameter-reduced forward model in the Gaussian likelihood case (cf. Section \ref{sec4}), the forward model $x\mapsto G(x)$ is replaced by $z\mapsto G(T(z))$. This way, the matrix $H^G$ should be replaced by
\begin{equation*}\label{eq:defHyG_new}
 H^G_z = \int_{\R^d} \nabla T(z)^\top \nabla G(T(z))^\top \Sigma_\text{obs}^{-1} \nabla G(T(z)) \nabla T(z)\, \mu(z)\d z.
\end{equation*}
Using either of these matrices, we obtain a projector $P_r$ to reduce the dimension in the variable $z=z_r+z_\perp$, where $z_r=P_r z$ and $z_\perp = (I_d-P_r)z$. 
In term of the original variable $x$, the dimension reduction method allows one to identify $x_r= T ( P_r T^{-1}(x) )$ with the observed data, while $x_\perp=T ( (I_d-P_r) T^{-1}(x) )$ is informed by the prior only.
Since $x_r$ and $x_\perp$ are nonlinear with respect to $x$, the resulting method can be interpreted as a \emph{nonlinear dimension reduction} method.

\section{Example 1: elliptic PDE} \label{sec:numerics_elliptic}

We first validate our methods using an inverse problem of identifying the coefficient of a two-dimensional elliptic PDE from point observations of its solution.

\subsection{Problem setup}
Consider the problem domain $\Omega = [0, 1]\times [0, 1]$, with boundary $\partial \Omega$. We denote the spatial coordinate by $s = (s_1, s_2) \in \Omega$.
We model the steady state potential solution field $p(s)$ for a given conductivity field $\kappa(s)$ and forcing function $f(s)$ using the Poisson's equation
\begin{equation}\label{eq:elliptic}
- \nabla \cdot \left( \kappa(s) \nabla p(s) \right) = f(s), \quad s \in \Omega. 
\end{equation}
Let $\partial \Omega_\tx{n} = \{ s \in \partial \Omega \,|\, s_2 = 0\}  \cup  \{ s \in \partial \Omega \,|\, s_2 = 1\}$ denote the top and bottom boundaries, and $\partial \Omega_\tx{d} = \{ s \in \partial \Omega \,|\, s_1 = 0\} \cup \{ s \in \partial \Omega \,|\, s_1 = 1\}$ denote the left and right boundaries. 
We impose the mixed boundary condition:
\[
p(s) = 0, \forall s \in \partial \Omega_\tx{d}, \quad \tx{and} \quad (\kappa(s) \nabla p(s) ) \cdot \vec{n}(s) = 0, \forall x \in \partial \Omega_\tx{n},
\] 
and let the forcing function take the form 
\[
f(s, t) = c\,\Big( \exp\big(-\frac{1}{2 r^2} \| s - a\|^2 \big) - \exp\big(-\frac{1}{2 r^2} \| s - b\|^2 \big) \Big), \forall t \geq 0,
\]
with $r = 0.05$, which is the superposition of two Gaussian-shaped sink/source terms centered at $a = (0.5, 0.5)$ and $b = (2.5, 0.5)$, scaled by a constant $c = 6\times 10^{-4}$.
The conductivity field $\kappa(s)$ is endowed with a log-normal prior. That is, letting $x(s) = \log \kappa(s)$, the Gaussian process prior for $x(s)$ is defined by
the stochastic PDE (see \cite{lindgren2011explicit} and references therein):
\begin{equation}\label{eq:heat_prior}
 - \triangle x(s) + \gamma x(s) = \mathcal{W}(s), \quad s\in\Omega,
\end{equation}
where $\triangle $ is the Laplace operator and $\mathcal{W}(s)$ is the white noise process. We impose a no-flux boundary condition on the above SPDE and set $\gamma = 10$.
Equations \eqref{eq:elliptic} and \eqref{eq:heat_prior} are solved using the finite element method with bilinear basis functions. A mesh with $80 \times 80$ elements is used in this example. This leads to $n = 6400$ dimensional discretised parameters.

\begin{figure}[h!]
\centering
\includegraphics[trim = 0em 0em 0em 0em, clip, width = 0.7\textwidth]{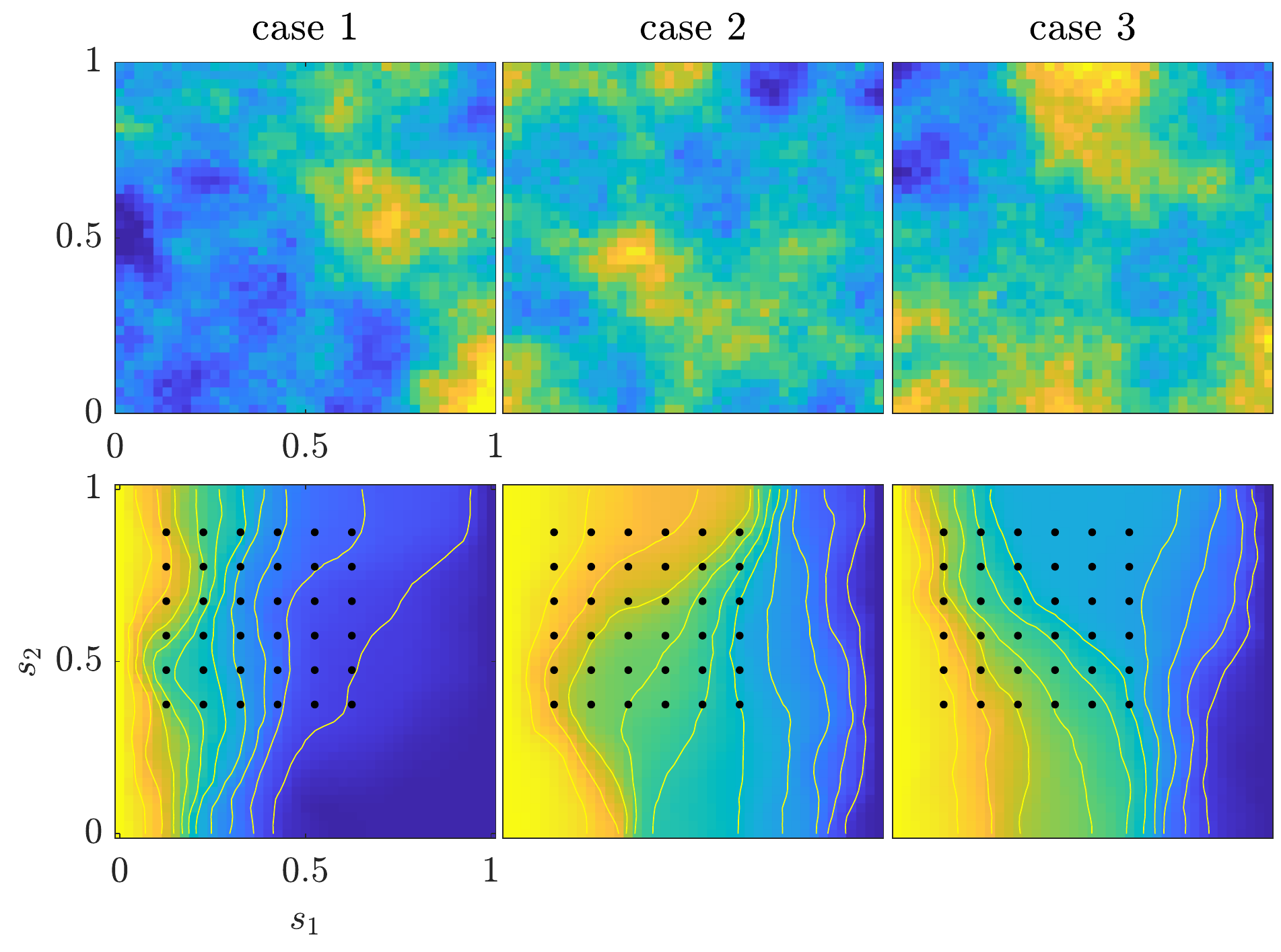}\vspace{-0.5em}
\caption{Setup of three test cases for the elliptic PDE example. The observation locations are shown as dots. Each column represent a test case, in which the top row shows the true conductivity fields and the bottom row shows the corresponding potential field.}
\label{fig:elliptic_setup}
\end{figure}

We generate three ``true'' conductivity fields from the prior distribution and use them to simulate synthetic observed data sets. 
The true conductivity fields and the simulated potential fields are shown in Figure \ref{fig:elliptic_setup}.
Observations of the potential fields are measured at the $m = 36$ discrete locations shown as black dots in Figure \ref{fig:elliptic_setup}. 
We set the standard derivation of the observation noise to $\sigma = 0.0415$, which corresponds to a signal-to-noise ratio of about $20$.

\subsection{Low-dimensional posterior approximations}

\begin{figure}[h]
 \centering
 \includegraphics[trim=2cm 0cm 2cm 0cm, width = 0.8\textwidth]{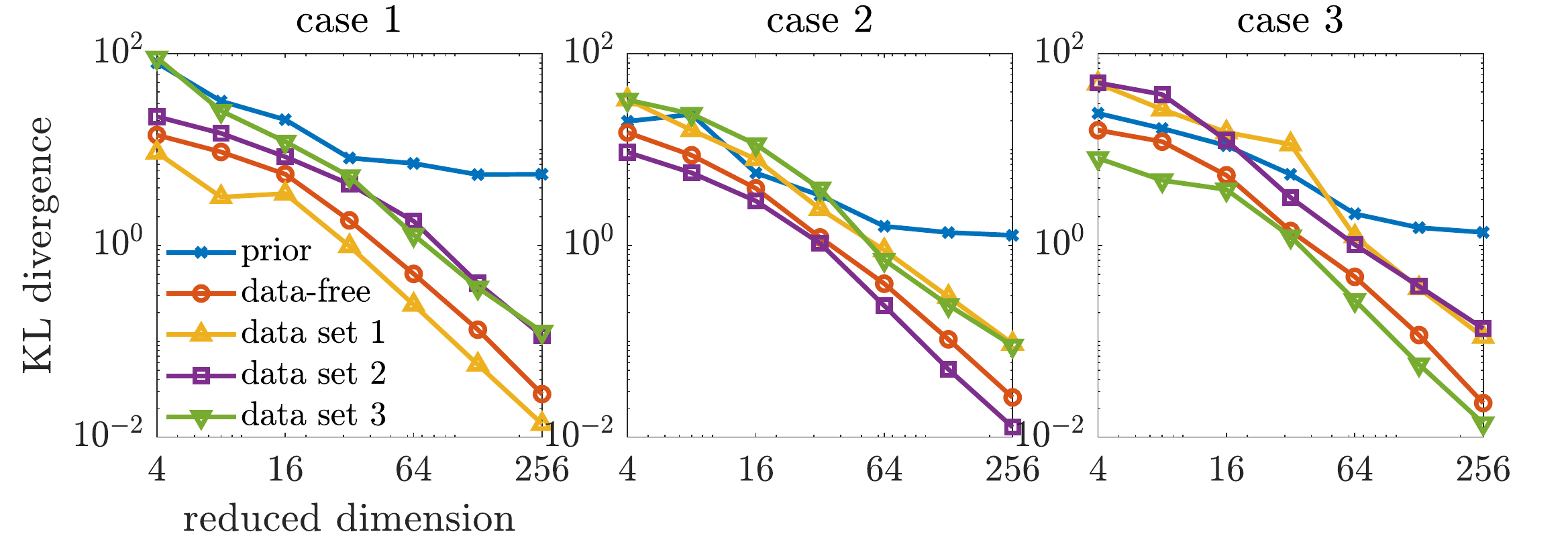}\\
 \includegraphics[trim=2cm 0cm 2cm 0cm, width = 0.8\textwidth]{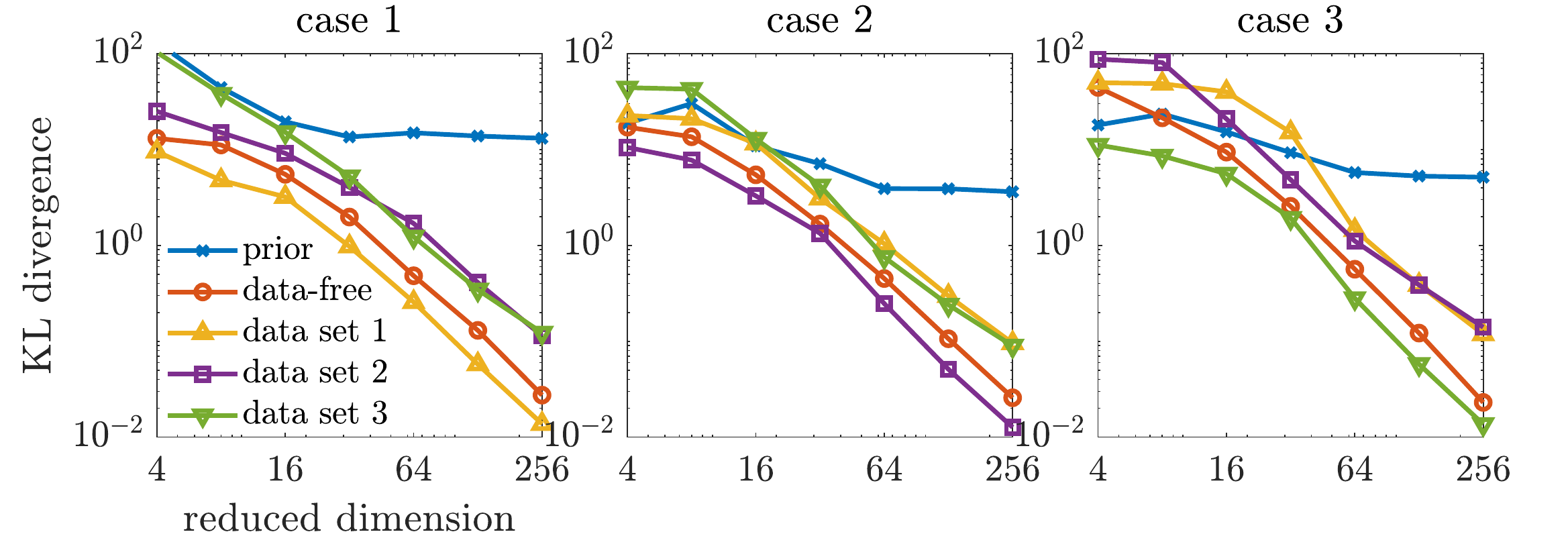}
 \caption{Elliptic PDE example. KL divergence of the full posterior densities from the approximate posterior densities defined by projectors with various ranks. Each column represents posteriors conditioned on a given data set. Top row: approximate posteriors are defined by the optimal parameter-reduced likelihood. Bottom: approximate posteriors are defined by the optimal parameter-reduced forward model. }\label{fig:elliptic_KLD}
\end{figure}

We first compare the approximate posterior densities defined by the data-free dimension reduction with that of the data-dependent dimension reduction and that of the truncated Karhunen--Lo\'{e}ve expansion of the prior.  
We build five sets of projectors: the data-free projectors as detailed in Section \ref{sec:DataFreeDR}, three sets of data-dependent projectors (see Section \ref{sec:DataDependentDR}) that correspond to three synthetic data sets, and projectors defined by leading eigenvectors of the prior covariance (\emph{i.e.} the truncated Karhunen--Lo\'{e}ve, referred to as the ``prior-based projector'' from hereinafter). 
For each of the data sets, the corresponding data-dependent projectors are constructed using the adaptive MCMC algorithm of \cite{zahm2018certified}.
Each set consists of projectors with ranks $r = 2^{2}, 2^{4}, \ldots, 2^{8}$.
For each projector, we compute the KL divergences of the full posteriors from the approximated posterior densities defined by the optimal parameter-reduced likelihood \eqref{eq:Loptimal}. The results are shown in the top row of Figure \ref{fig:elliptic_KLD}.
Using the same set of projectors, we also compare the KL divergences of the full posteriors from the resulting approximated posterior densities defined by the optimal parameter-reduced forward model \eqref{eq:Goptimal}. The results are shown in the bottom row of Figure \ref{fig:elliptic_KLD}.

In these experiments, we estimate the KL divergence using Monte Carlo integration with $N$ posterior samples, which yields 
\[
D_{KL}( \pi^y_\text{pos} \| \tilde{\pi}^y_\text{pos} ) \approx \frac{1}{N}  \sum_{i = 1}^N \big( \log \mathcal{L}^y(x^{(i)})- \log \tilde{\mathcal{L}}^y(x^{(i)}) \big) + \log\Big(\frac{1}{N}\sum_{i = 1}^N \exp\big( \log \tilde{\mathcal{L}}^y(x^{(i)}) - \log \mathcal{L}^y(x^{(i)}) \big)\Big),
\]
where the second sample average accounts for the ratio between normalizing constants. For approximations that are close to the full posterior, using a reasonable number of (independent) posterior samples, e.g., $10^5$ used here, make the standard deviations of the estimated KL divergence insignificant compared with the mean estimates in our numerical examples.

We observe that the optimal parameter-reduced likelihood and the optimal parameter-reduced forward model result in approximate posteriors with similar accuracy. 
For sufficiently large ranks ($r \geq 8$), the most accurate approximate posterior densities are obtained by the data-dependent projectors of the corresponding data set, followed by those obtained by the data-free projectors.
We also observe that, for each data set, the data-dependent projectors constructed using other data sets result in less accurate approximations. 
By allowing a marginal loss of accuracy compared to the data-dependent construction, the data-free construction bypasses the computationally costly online dimension reduction process for every new data set. 
Compared with the prior-based dimension reduction, which is also an offline method, the data-free construction offers significantly more accurate approximations in this example.

\begin{figure}[h]
 \centering
 \includegraphics[trim=2.5cm 1cm 2cm 0cm, width = 0.53\textwidth]{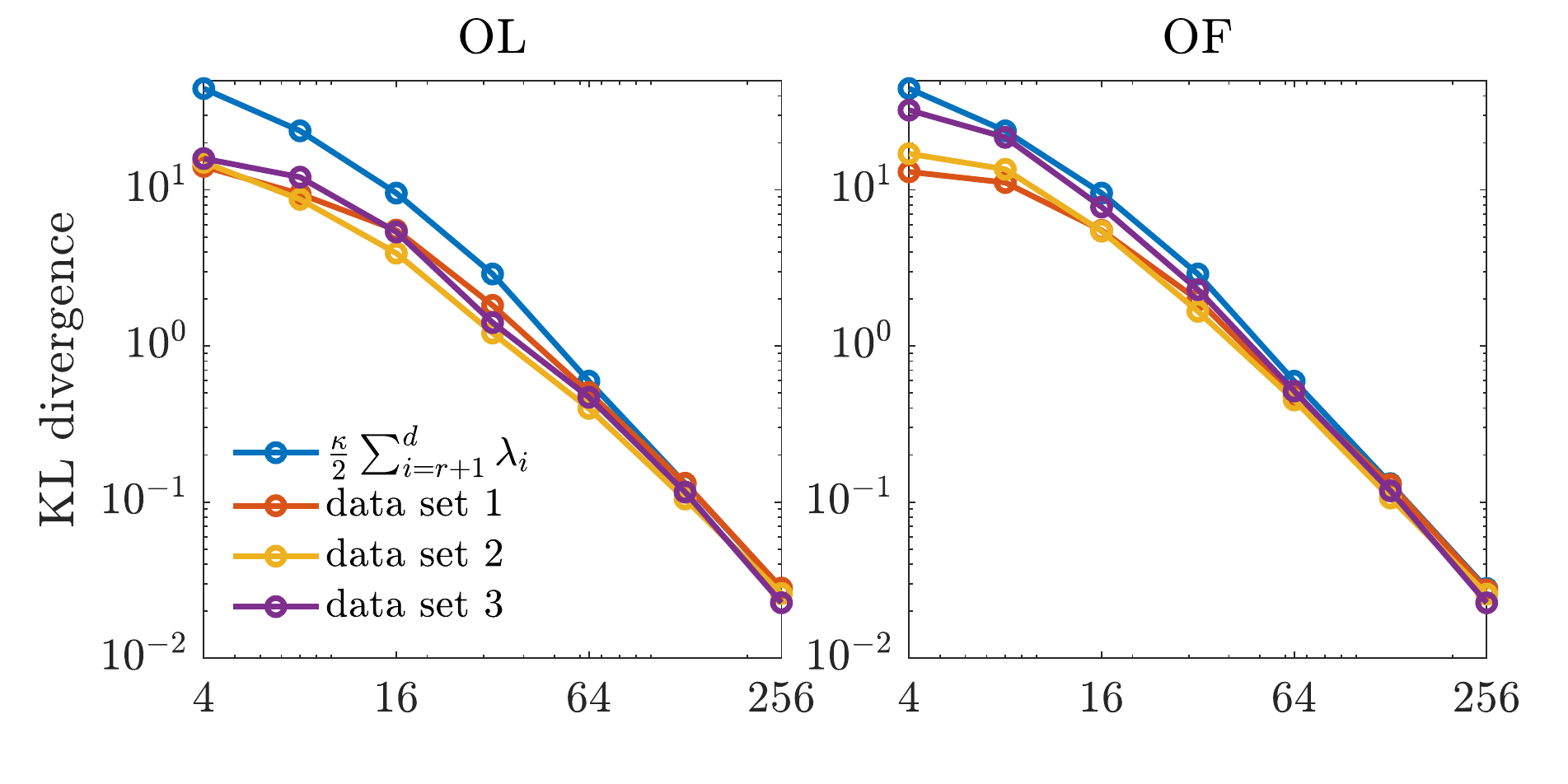}
 \caption{Elliptic PDE example. The bounds in \eqref{eq:KLBoundExpectation} and \eqref{eq:KLBoundExpectation_BIS} (with $\kappa =1$) are compared to KL divergence of the full posterior densities from the approximate posterior densities defined by the data-free projectors with various ranks. Left: approximate posteriors are defined by the optimal parameter-reduced likelihood. Right: approximate posteriors are defined by the optimal parameter-reduced forward model.}\label{fig:ellptic_bound}
\end{figure}

For each of the data sets, we also compare the errors of the approximate posterior densities with the bounds defined in \eqref{eq:KLBoundExpectation} and \eqref{eq:KLBoundExpectation_BIS}. Note that the right hand sides of \eqref{eq:KLBoundExpectation} and \eqref{eq:KLBoundExpectation_BIS} are the same up to the constant $\kappa$ in this example. We plot the errors and the bounds (with $\kappa =1$ for Gaussian prior) in Figure \ref{fig:ellptic_bound}, in which all approximate posterior densities are defined by the data-free projectors. 
In this example, we observe that the errors of the approximate posterior densities follow the same trend as their corresponding error bounds. Note that both \eqref{eq:KLBoundExpectation} and \eqref{eq:KLBoundExpectation_BIS} give upper bounds on the expected KL divergence, and thus they may not bound the KL divergence for a realization of the data.

\subsection{Subspace accelerated sampling}

\begin{table}[h]
\caption{Acronyms of various inference algorithms used in numerical comparisons.}
\label{table:acronyms}
\centering
\begin{tabular}{l | p{0.75\textwidth}}\hline
OL & approximate inference using Algorithm \ref{alg:MCMC_approx} and the optimal parameter-reduced likelihood function in \eqref{eq:Loptimal_sample}\\
PM & exact inference using the pseudo-marginal method (Algorithms \ref{alg:PseudoMarginalMCMC} and \ref{alg:MCMC_for_exact_inference})\\ \hline
OF & approximate inference using Algorithm \ref{alg:MCMC_approx} and the optimal parameter-reduced forward model in \eqref{eq:Goptimal_sample}\\\hline
DA & exact inference using the delayed acceptance algorithm (Algorithm \ref{alg:DA_MCMC}) with the approximation defined by the parameter-reduced forward model in \eqref{eq:Goptimal_sample}\\ \hline
H-MALA & exact inference using the Hessian preconditioned Langevin MCMC \cite{petra2014computational}\\\hline
PCN & exact inference using the preconditioned Crank--Nicolson MCMC \cite{beskos2008mcmc,cotter2013mcmc}\\\hline
\end{tabular}
\end{table}

We demonstrate the sampling performance of various approximate and exact inference algorithms introduced in Sections \ref{sec5} and \ref{sec6} using the posterior density conditioned on the first data set. All the methods used in the comparison and their acronyms are summarized in Table \ref{table:acronyms}. 

We use the Hessian-preconditioned Metropolis-Adjusted Langevin Algorithm (H-MALA) and the preconditioned Crank--Nicolson (PCN) MCMC as reference MCMC methods for sampling the full-dimensional posterior. Since H-MALA uses the low-rank decomposition of the Hessian matrix of the logarithm of the posterior density computed at the \emph{maximum a posteriori} point to precondition MCMC, it can also be viewed as a data-dependent subspace-accelerated method. We refer to \cite{cui2016dimension,petra2014computational} for a detailed discussion. 
In order to make a fair comparison with H-MALA, the MCMC algorithm we use on our data-free informed subspace is based on a Langevin proposal preconditioned by the same Hessian matrix used by H-MALA projected onto the data-free informed subspaces.

\begin{figure}[ht]
 \centering
 \includegraphics[trim=2.5cm 0cm 2cm 0cm, width = 0.55\textwidth]{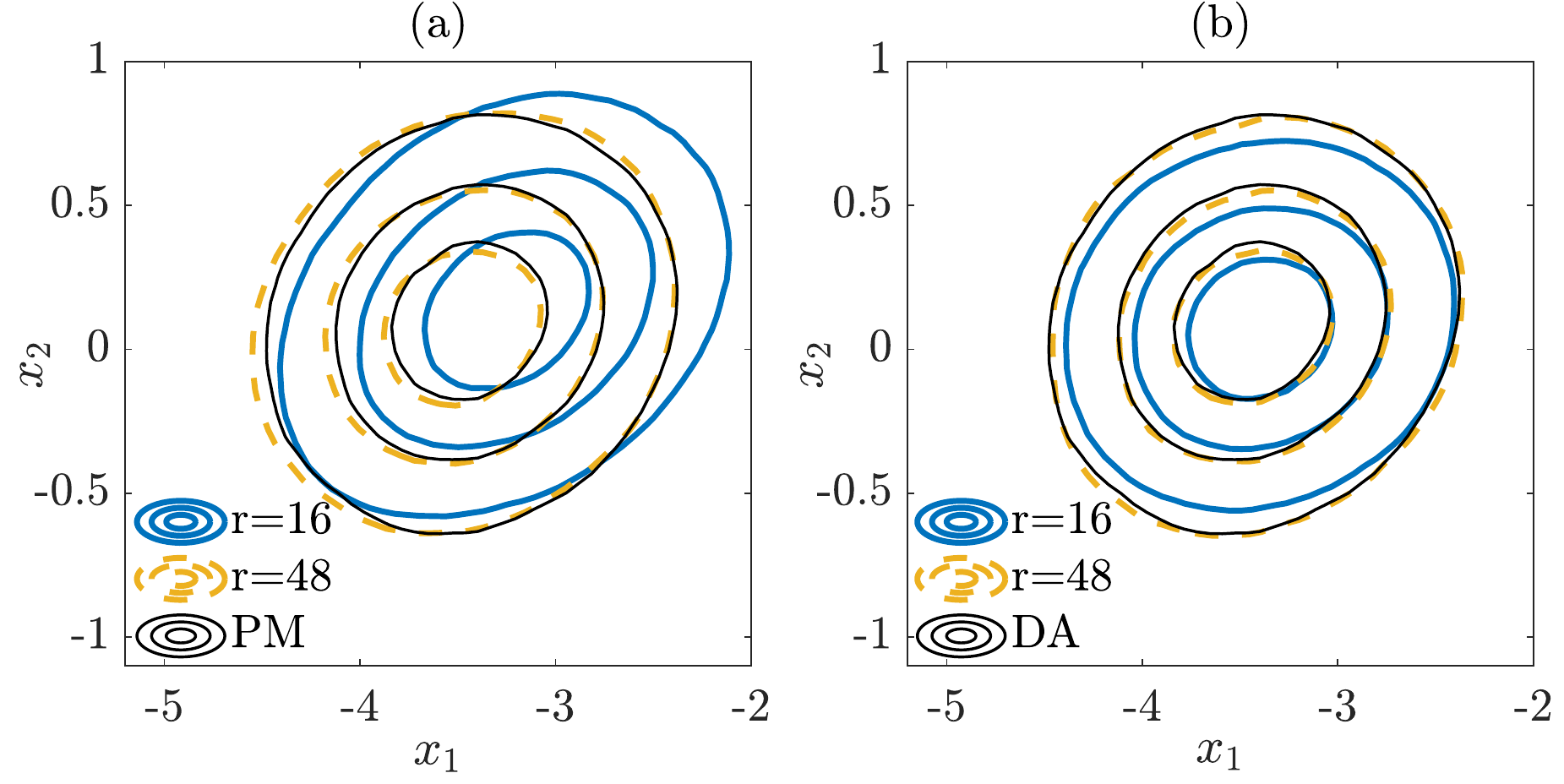}
 \caption{Elliptic PDE example. Contours of the marginal posterior densities computed by various inference algorithms using the data-free projector.  (a): OL (with $r = 16$ and $r=48$) and PM (with $r=48$). (b): OF ($r = 16$ and $r=48$) and DA (with $r = 48$). Here $(x_1, x_2)$ represent the directions spanned by the first two data-free LIS basis vectors.}\label{fig:contour_elliptic}
\end{figure}

In Figure \ref{fig:contour_elliptic}, the contours of the marginal posterior densities (marginalized onto the first two data-free LIS basis vectors) produced by approximate inference methods (with $r = 16$ and $r=48$) are compared with those produced by their exact inference modifications (with $r=48$).  
We can observe that the results produced by approximate inference methods approach those of their modifications as the rank of informed subspace increases.

To measure the efficiency of various MCMC methods, we use the average integrated autocorrelation times (IACTs) of parameters
\[
\tau = \frac{1}{d}\sum_{i = 1}^d {\rm IACT}(x_i),
\] 
where ${\rm IACT}(x_i)$ is the IACT of the $i$-th component of $x$. See \cite[Section 12.7]{liu2001monte} for the definition of IACT.
The data-free projectors with different ranks $r$ and two different sample sizes $N=2$ and $N=5$ are used in this experiment. Here H-MALA and PCN are used as base cases to benchmark those MCMC methods accelerated by the informed subspace. 
All the methods are simulated for $5\times10^5$ iterations and repeated $10$ times to report the mean and the standard deviation of $\tau$. The initial state of all the simulations are randomly selected from a pre-computed Markov chain of posterior samples to avoid burn-in. 
The results are reported in Table \ref{table:iact_elliptic}.

For the approximate inference methods (OL and OF), the average IACTs consistently increase with the rank of the projectors, as the sampling performance of the Langevin proposal is expected to decay with the underlying parameter dimension. 
Both OL and OF produce significantly smaller IACTs compared with the full-dimensional H-MALA.

\begin{table}[h]
\caption{Elliptic PDE example. Average IACTs of parameters computed by various inference algorithms applied to the posterior conditioned on data set 1. Here the symbol - indicates poorly mixing Markov chains that do not have reliable estimate of the IACTs. All the data reported here are in the form of mean$\pm$standard derivation.}
\label{table:iact_elliptic}
\centering
\begin{tabular}{l@{\hspace{1\tabcolsep}}l| l@{\hspace{1\tabcolsep}}l@{\hspace{1\tabcolsep}}l @{\hspace{0.5\tabcolsep}}|@{\hspace{0.5\tabcolsep}} l@{\hspace{1\tabcolsep}}l@{\hspace{1\tabcolsep}}l | l @{\hspace{1\tabcolsep}}l}\hline
    && \multicolumn{2}{c}{IACT} & & \multicolumn{2}{c}{IACT} & & IACT \\ \cline{3-4}\cline{6-7}\cline{9-10}
    && \;OL & \;PM & \hspace{-1.2em}$\sqrt{\textrm{var}}[\log\widetilde{\mathcal{L}}_N^y]$\hspace{-0.3em} & \;OF & \;DA &  \;$\E[\beta]$ & HMALA & \;\;PCN \\ \hline
\multirow{5}{*}{\begin{sideways}$N=2$\end{sideways}} 
  & $r = 16$ & 18.9$\pm$1.5 & 163$\pm$29  & 4.45$\pm$0.20 & 19.7$\pm$1.2 & -   & $<$0.1 & \multirow{10}{*}{164$\pm$17} & \multirow{10}{*}{1303$\pm$139}\\
  & $r = 24$ & 34.9$\pm$1.1 & 106$\pm$13   & 2.65$\pm$0.19 & 35.7$\pm$2.1 & -   & $<$0.1 & \\
  & $r = 32$ & 52.6$\pm$3.0 & 91.8$\pm$5.3 & 1.80$\pm$0.10 & 57.1$\pm$3.1 & 208$\pm$39 & 0.31$\pm$0.02 & \\
  & $r = 40$ & 59.4$\pm$2.4 & 91.6$\pm$6.1 & 0.93$\pm$0.03 & 63.0$\pm$2.1 & 208$\pm$26 & 0.36$\pm$0.02 & \\
  & $r = 48$ & 60.7$\pm$2.4 & 83.8$\pm$5.6 & 0.69$\pm$0.02 & 66.9$\pm$4.3 & 146$\pm$10 & 0.46$\pm$0.01 & \\ \cline{1-8}
\multirow{5}{*}{\begin{sideways}$N=5$\end{sideways}} 
  & $r = 16$ & 18.7$\pm$1.0 & 102$\pm$8.2 & 2.28$\pm$0.10 & 19.3$\pm$1.3 & -   & $<$0.1 & \\
  & $r = 24$ & 32.7$\pm$1.7 & 72.6$\pm$4.3 & 1.38$\pm$0.05 & 37.8$\pm$2.5 & 255$\pm$36 & 0.19$\pm$0.02 & \\
  & $r = 32$ & 48.8$\pm$1.2 & 71.6$\pm$3.0 & 0.97$\pm$0.06 & 55.8$\pm$1.1 & 214$\pm$38 & 0.31$\pm$0.01 & \\
  & $r = 40$ & 55.2$\pm$2.1 & 67.4$\pm$3.4 & 0.55$\pm$0.03 & 61.7$\pm$2.8 & 173$\pm$21 & 0.39$\pm$0.01 & \\
  & $r = 48$ & 56.0$\pm$3.3 & 64.9$\pm$3.2 & 0.41$\pm$0.02 & 69.9$\pm$3.5 & 148$\pm$26 & 0.47$\pm$0.01 & \\ \hline
\end{tabular}
\end{table}

Compared to the OL method, the PM method (the exact inference counterpart for OL) has a different behavior. Here we recall that the sample-averaged parameter-reduced likelihood, $\widetilde{\mathcal{L}}_N^y$, in the PM method is a random estimator, whereas $\widetilde{\mathcal{L}}_N^y$ in the OL method is deterministic because of the usage of prescribed samples. 
The standard deviation of the logarithm of $\widetilde{\mathcal{L}}_N^y$ in Table \ref{table:iact_elliptic} confirms that low-rank projectors have rather large Monte Carlo errors as the approximation accuracy is controlled by the rank truncation (cf. \eqref{eq:KLBoundExpectation}). 
The exactness of the PM method comes at the cost of Monte Carlo error, which is controlled by the sample size $N$ and the rank of the projector. 
We observe that increasing either the rank or the sample size can narrow the gap between the IACTs produced by PM and its OL counterpart. 
This experiment clearly suggests that PM needs to balance the sample size $N$ and the rank of the projector to achieve the optimal performance.

Compared to the OF method, the DA method (the exact inference counterpart for OF) produces the largest IACTs among all subspace inference methods. This result is not surprising, as the second stage acceptance/rejection of DA necessarily deteriorates the statistical performance \cite{christen2005markov}. In Table \ref{table:iact_elliptic}, we observe that the second stage acceptance rates, $\E[\beta]$, increase with more accurate approximations obtained with higher projector ranks and larger sample sizes. As the result, the gaps between the IACTs produced by OF and DA are smaller for higher projector ranks and larger sample sizes.

Overall, approximate inference methods have better statistical performance compared to other methods in this example (cf. Table \ref{table:iact_elliptic}) and can obtain reasonably accurate results as shown in Figures \ref{fig:elliptic_KLD} and Figure \ref{fig:contour_elliptic}. With the additional cost that comes in the form of either Monte Carlo error (PM) or the second stage acceptance/rejection (DA), the exact inference modifications produce Markov chains with larger IACTs. 
Among all the exact inference methods, PM produces smaller IACTs compared with the full-dimensional H-MALA, PCN, and DA. 

It is worth to mention that each iteration of the subspace MCMC method needs $N$ number of forward model simulations, whereas H-MALA requires only one forward model simulation per iteration. In this example, approximate inference methods (OL and OF) with $N=2$ still outperforms H-MALA in terms of IACTs per model evaluation. Exact inference methods, however, need more model evaluations than H-MALA to obtain the same number of effective samples (we will show in the subsection another example where H-MALA is outperformed by PM and DA).
Notice that the forward model evaluations in each iteration can be embarrassingly parallelized: with parallel computing resources available, the subspace MCMC methods can still be more efficient than H-MALA in terms of the effective sample size per wall-clock time.

\section{Example 2: PET with Poisson data}\label{sec:poisson_numerics}

The second example is a two dimensional PET imaging problem, where we aim to reconstruct the density of the object from integer-valued Poisson observed data.
We use here a Besov prior for which we access the coordinate selection technique and the prior normalization method presented in Sections \ref{sec_coor_select} and \ref{sec_normalization}.

\subsection{Problem setup}

\begin{figure}[!htbp]
\begin{subfigure}{0.38\textwidth}\centering
\centerline{\includegraphics[trim = 0em 0em 0em 0em, clip, width = 1\textwidth]{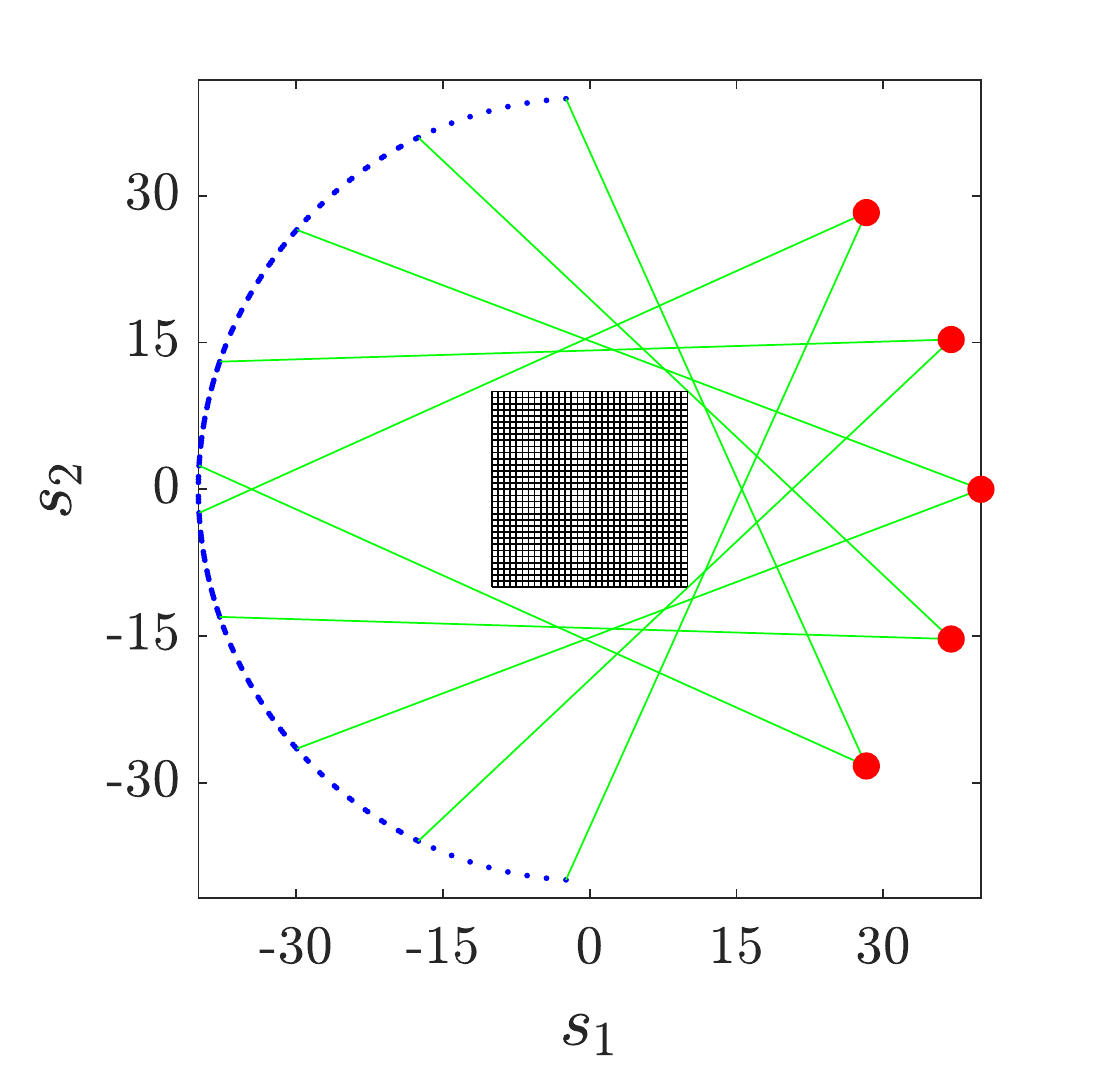}\vspace{-0.3em}}
\caption{Discretised domain of interest $\Omega = [-10, 10]^2$ (mesh), radiation sources (red dots), and detectors (blue dots). }\label{fig:pet_setup}
\end{subfigure}\hfill
\begin{subfigure}{0.57\textwidth}
\hspace{-0.3em}\includegraphics[trim = 0em 1em 0em 0em, clip, width = 1\textwidth]{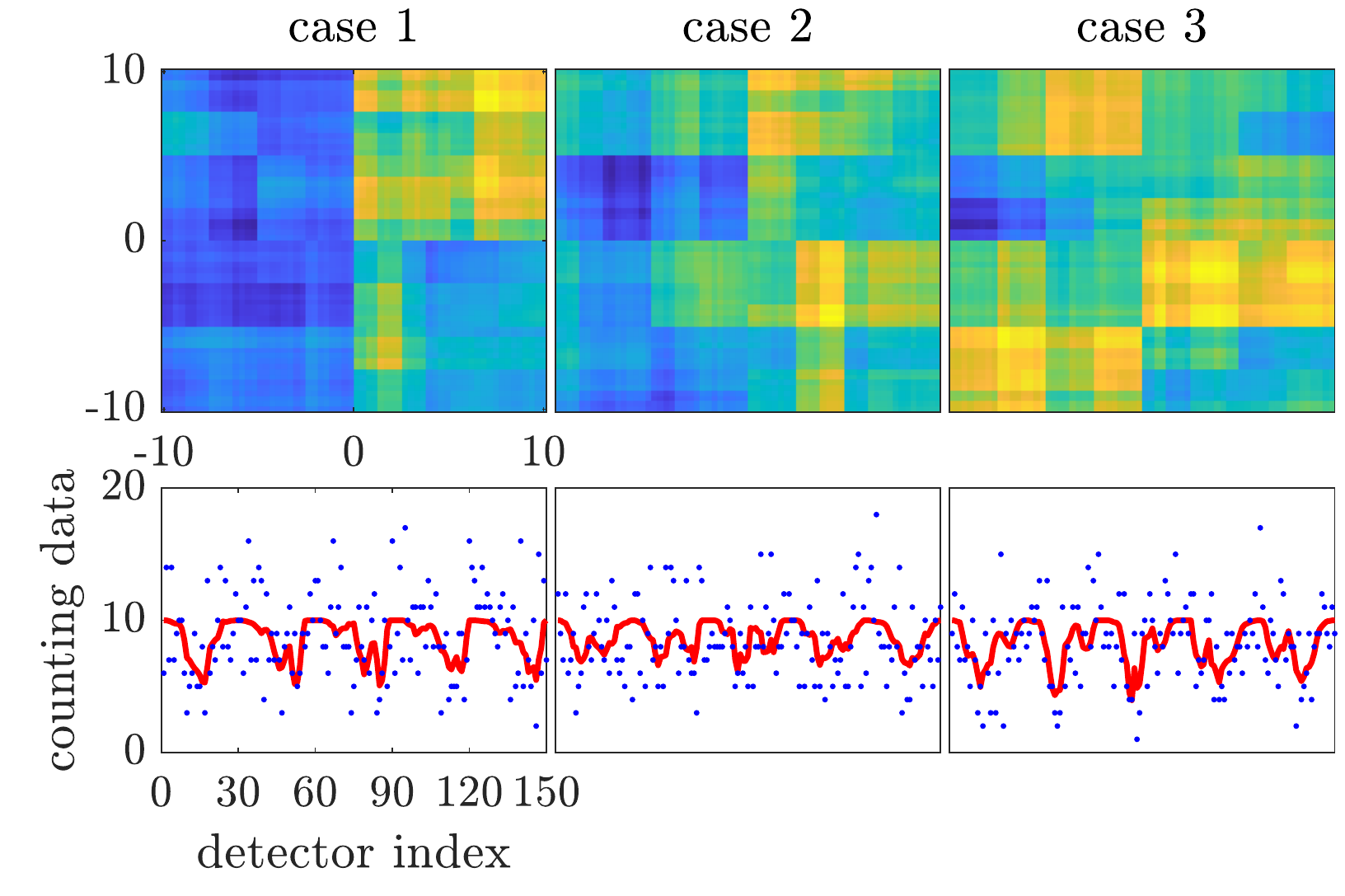}\vspace{-0.5em}
\caption{Top row: three density functions generated from the prior. Bottom row: corresponding intensity function (solids lines) and measured counting data sets (dots).}\label{fig:pet_setup_cases}
\end{subfigure}
\caption{The PET setup and three test cases.}
\end{figure}

In PET imaging, the goal is to identify an object of interest located inside a domain $\Omega$ subjected to gamma rays.
The rays travel through $\Omega$ from multiple sources and the detectors count the number of incident photons (thus the data are integer-valued), see Figure \ref{fig:pet_setup}.
The object of interest is described by its density of mass 
which is represented by $s \mapsto \exp(f(s))$, where $f:\Omega\rightarrow \R$ follows a Besov-$\mathcal{B}^2_{11}$ prior with the Haar wavelet, see Section \ref{sec_besov_2}.
The change of intensity of a gamma ray along the path, $\ell_i(s), s \in \Omega$, can be modelled using Beer's law:
\begin{gather}\label{eq:beers}
I_{d, i}  = I_{s,i} \exp \Big( - \int_{\ell_{i}(s)} \exp \big( f(s) \big)  d s \Big),
\end{gather}
where $I_{d,i} \in \R_{\geq 0}$ and $I_{s,i} \in \R_{\geq 0}$ are the intensities at the detector and at the source, respectively.
We assume that all the gamma ray sources have the same intensity, $I_{s,i} = I_s$ for $i = 1, \ldots, m$.

In this example, the domain $\Omega$ is discretized into a regular grid with $d$ cells and the logarithm of the density is assumed to be piecewise constant. 
This yields the discretized parameter $x \in \R^d$. %
The line integrals in \eqref{eq:beers} are approximated by
\begin{equation*}
\int_{\ell_{i}(s)} \exp ( f(s) )  d s \approx \sum_{j = 1}^{n} A_{ij} \, \exp(f_j),
\end{equation*}
where $A_{ij} \in \R_{\geq 0}$ is the length of the intersection between line $\ell_{i}$ and cell $j$, and $\exp(f_j)$ is the discretized density in cell $j$.
By discretizing the wavelet basis on the same grid and following the parametrization discussed in Section \ref{sec_besov}, we can write
\[
f = B x,
\]
where $B \in \R^{d \times d}$ consists of discretized basis functions and $x$ consists of associated coefficients. 
In this setting, $x$ follows a product-form Laplace distribution given by \eqref{eq:PriorWithIndepCompo} with $p=1$ and with the scale parameter arbitrarily set to $\gamma=1$. 
Suppose we have a total of $m$ number of gamma ray paths and the corresponding matrix $A \in \R^{m \times d}$, the forward model $G: \R^{d} \rightarrow \R^{m}$ is defined as
\begin{gather}
G(x) = I_{s}\,\exp( - A \, \exp( Bx) ).
\label{eq:forward_xray}
\end{gather}
We consider a PET setup shown in Figure \ref{fig:pet_setup}: the problem domain $\Omega = [-10, 10]^2$ is discretised into a $d = 64\times 64$ regular grid, five radiation sources with intensity $I_s = 10$ are positioned with equal spaces on one side of a circle, spanning a $90^{\circ}$ angle, and each source sends a fan of 30 gamma rays that are measured by detectors. This leads to $m = 150$ observations. The model setup is based on the code of \cite{heikkinen2008statistical}.

We denote the observed data by $y \in \mathbb{N}^{m}$ where each element $y_i$ is associated with the $i$-th gamma ray in the model.
For the $i$-th gamma ray, recall that the intensity at the detector is computed by $G_i(x)$ for some input parameter $x$, and then the probability mass function of observing the counting data $Y_i = y_i$ is given by
\[
 \mathbb{P}(Y_i = y_i | x) = \frac{G_i(x)^{y_i}\, \exp(- G_i(x)) }{y_i !}. %
\]
Suppose we can observe the counting data at all the detectors and assume the measurement processes are independent, we can write the likelihood function with the complete data as
\begin{align}
\mathcal{L}^y(x) = \prod_{i = 1}^{m} \mathbb{P}(Y_i = y_i | x) = \prod_{i = 1}^{m} \frac{G_i(x)^{y_i}\, \exp(- G_i(x)) }{y_i !}.
\end{align}
As shown in \ref{sec:fisher}, the Fisher information matrix of the above likelihood function takes the form
\begin{align}
\mathcal{I}(x) = \nabla G(x)^\top M(x) \nabla G(x),
\end{align}
where $M(x)$ is a diagonal matrix with $M_{ii}(x) = G_i(x)^{-1}$ along its diagonal. 
We generate three ``true'' density functions from the prior distribution and use them to simulate synthetic data sets. 
The true density functions and the simulated data are shown in Figure \ref{fig:pet_setup_cases}.

\subsection{Numerical results using coordinate selection}

We first present the results obtained by applying the coordinate selection method (cf. Section \ref{sec_coor_select}). Similar to the first example, here we will first compare the accuracy of approximate posterior densities defined by various approaches and projectors, and then benchmark the performance of MCMC methods. We adopt the same setup and acronyms as in Example 1 and Table \ref{table:acronyms}. 

\begin{figure}[h]
 \centering
 \includegraphics[trim=2.5cm 0cm 2cm 0cm, width = 0.8\textwidth]{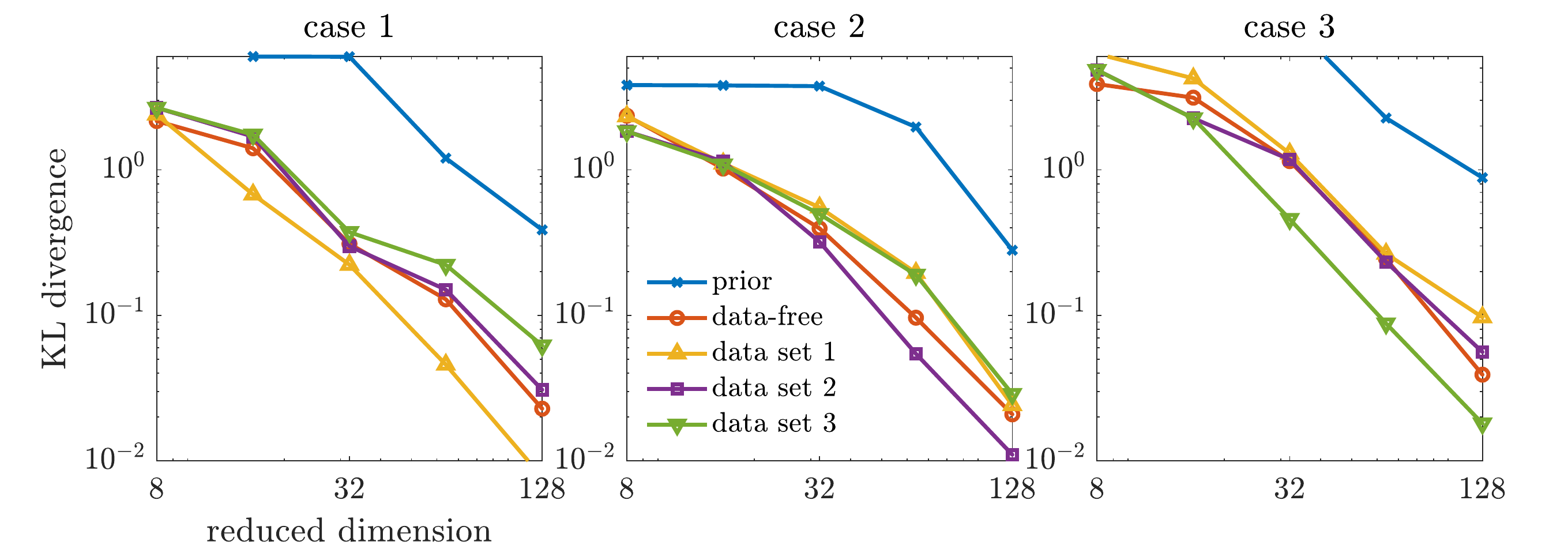}\\
 \includegraphics[trim=2.5cm 0cm 2cm 0cm, width = 0.8\textwidth]{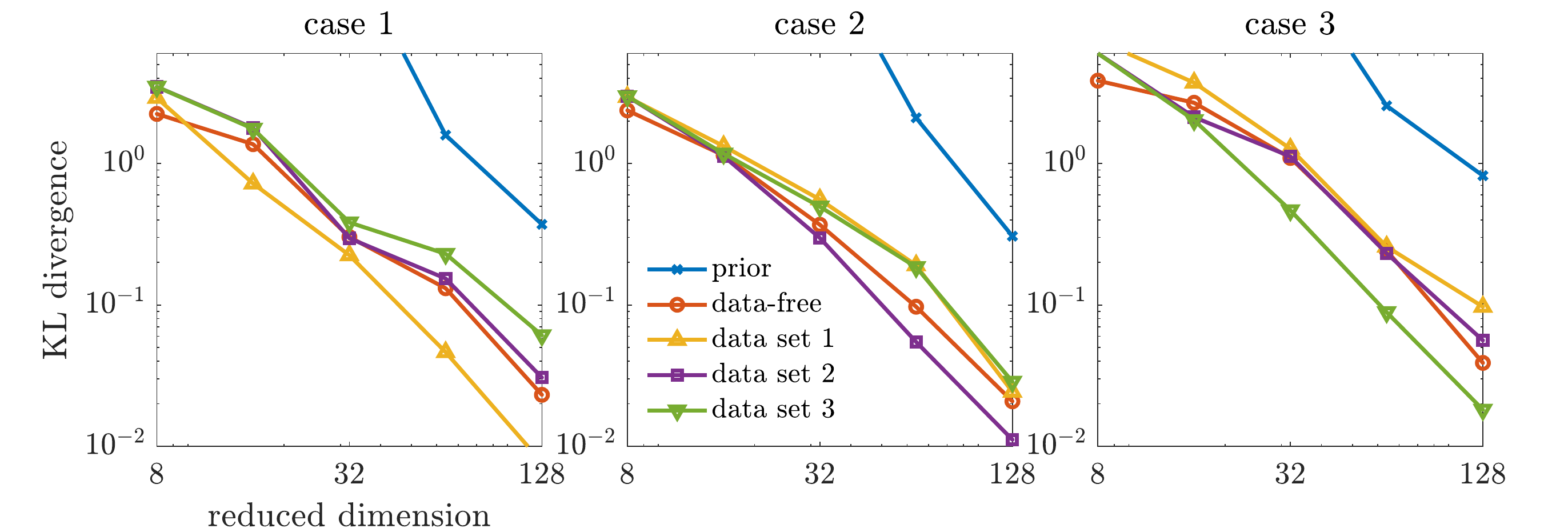}
 \caption{Same as Figure \ref{fig:elliptic_KLD}, but for the PET test case and where we used coordinate selection.}\label{fig:pet_KLD}
\end{figure}

For the approximate posterior densities, five sets of projectors built from selected coordinates, including the data-free projectors, three sets of data-dependent projectors (corresponding to three data sets), and the prior-based truncated wavelet basis, are considered. 
Each set consists of projectors with ranks $r = 2^{3}, 2^{4}, \ldots, 2^{7}$.
The KL divergences of the full posteriors from the approximated posterior densities defined by the optimal parameter-reduced likelihood \eqref{eq:Loptimal} are shown in the top row of Figure \ref{fig:pet_KLD}, while those of the optimal parameter-reduced forward model \eqref{eq:Goptimal} are shown in the bottom row of Figure \ref{fig:pet_KLD}.

In this example, we observe similar results as the results of the elliptic PDE example. 
The optimal parameter-reduced likelihood and the optimal parameter-reduced forward model result in approximate posteriors with similar accuracy. 
The most accurate approximate posterior densities are obtained by the data-dependent projectors of the corresponding data set, followed by those obtained by the data-free projectors. For each data set, the data-dependent projectors constructed using other data sets result in less accurate approximations in general. However, the accuracy gaps between the data-free projectors and the data-dependent projectors (using other data sets) are not as significant as the elliptic PDE example.
This can be caused by either the coordinate selection method or the rather large data size in this example.
Compared with the prior-based dimension reduction, which is also an offline method, the data-free construction offers significantly more accurate approximations in this example. 
Overall, the data-free dimension reduction provides reasonably accurate posterior approximations for the Poisson observation process considered here. 

\begin{figure}[h]
 \centering
 \includegraphics[trim=2.5cm 1cm 2cm 0cm, width = 0.53\textwidth]{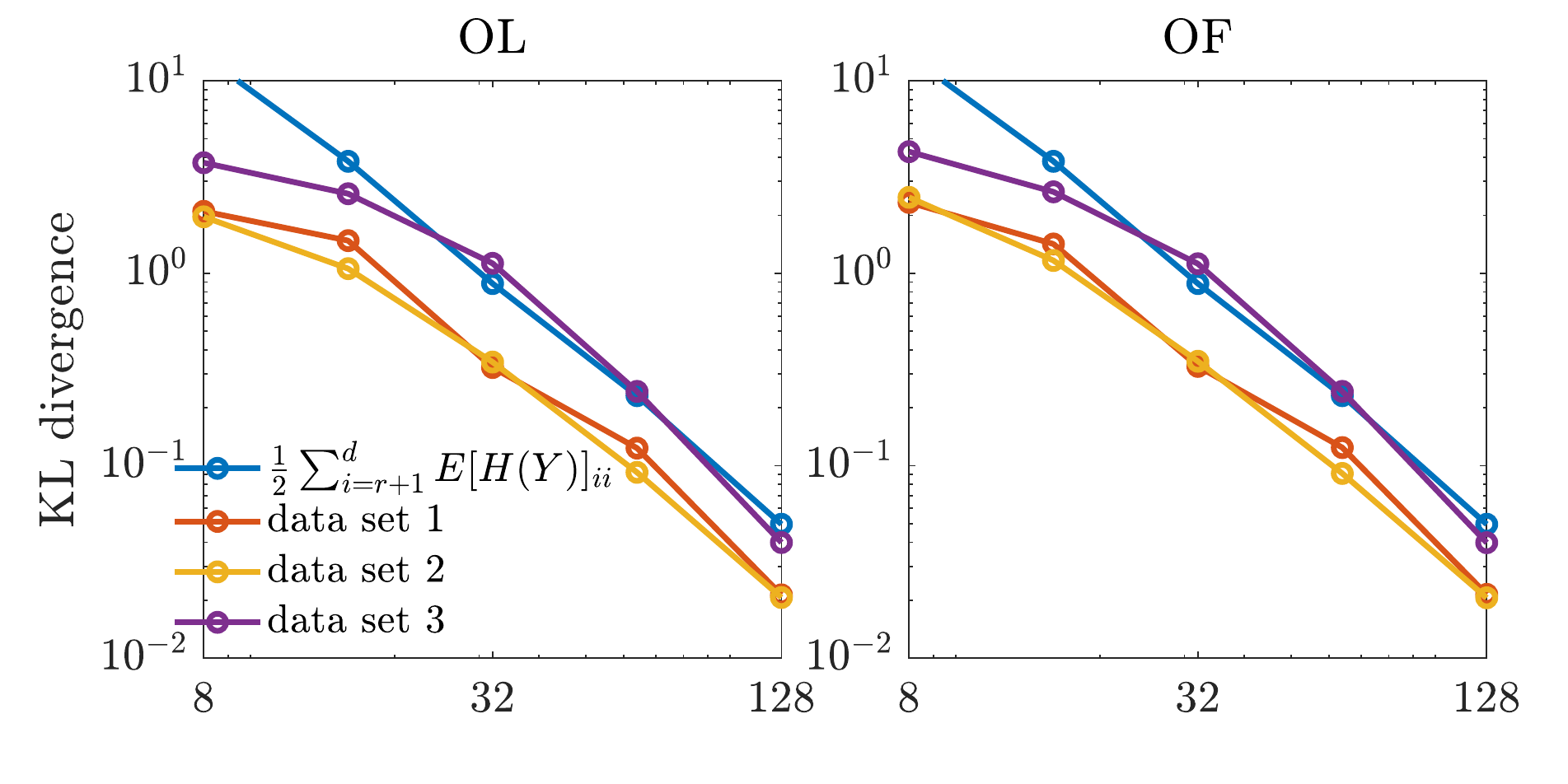}
 \caption{Same as Figure \ref{fig:ellptic_bound}, but for the PET test case and where we used coordinate selection.}\label{fig:pet_bound}
\end{figure}

Although it remains an open question if the bounds in \eqref{eq:KLBoundExpectation} and \eqref{eq:KLBoundExpectation_BIS} can be applied for Besov priors, we provide a comparison of the errors of the approximate posterior densities (defined by the data-free projectors) with the bounds. The results are shown in Figure \ref{fig:pet_bound} with $\kappa$ being replaced by $1$. 
Interesting, we still observe that the errors of the approximate posterior densities follow the same trend as their corresponding bounds.

\begin{figure}[h]
 \centering
 \includegraphics[trim=2.5cm 0cm 2cm 0cm, width = 0.6\textwidth]{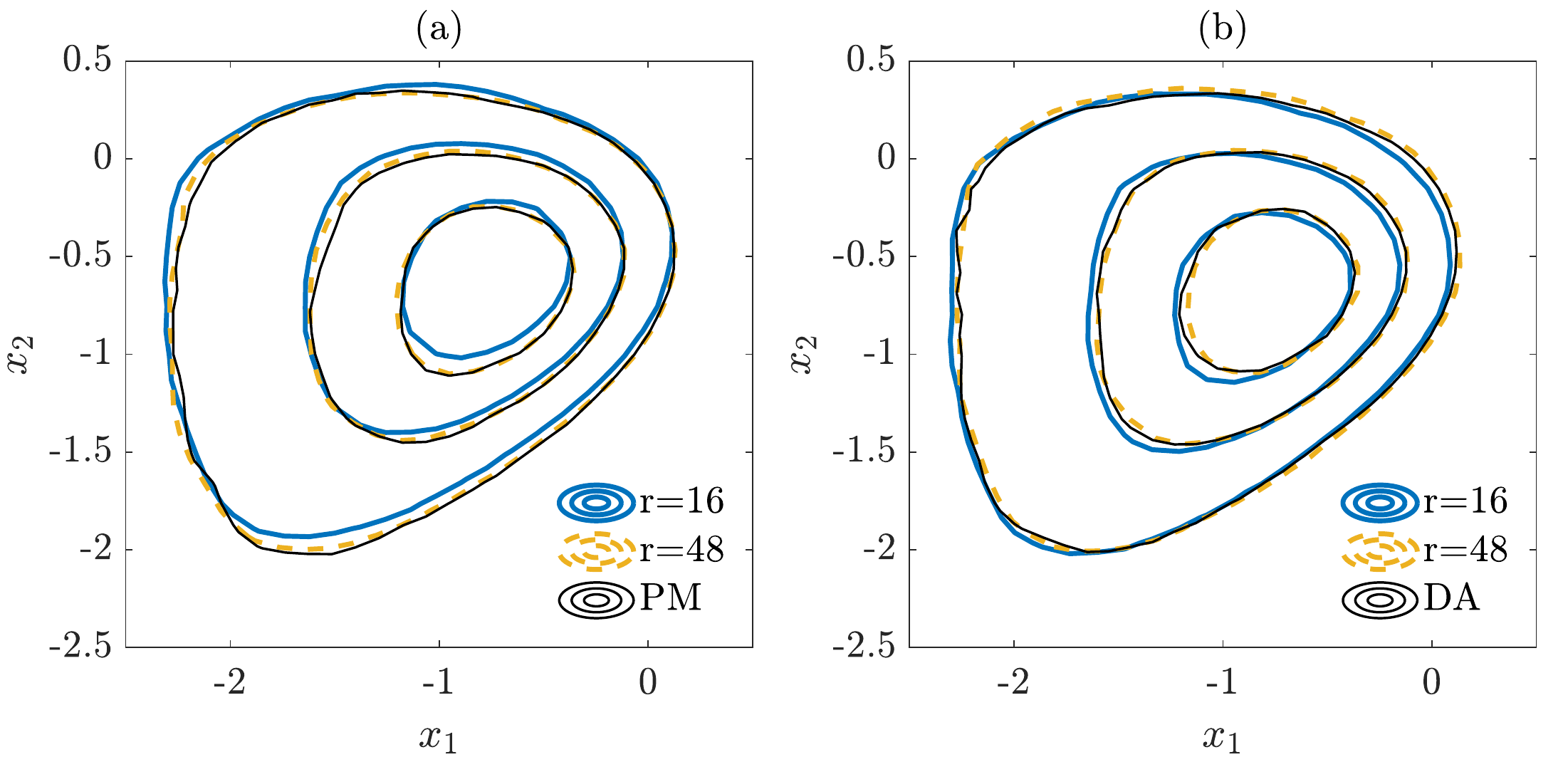}
 \caption{Same as Figure \ref{fig:contour_elliptic}, but for PET and data set 1. Here $r = 16, 48$ are used in OL and OF, and $r=48$ is used in PM and DA. Here $(x_1, x_2)$ represent the first two coordinates selected by the data-free method.}\label{fig:contour_pet}
\end{figure}

We then compare the performance of various subspace driven inference methods. In Figure \ref{fig:contour_pet}, the contours of the the marginal posterior densities produced by approximate inference methods (with $r = 16$ and $r=48$) are compared with those produced by their exact inference modifications (with $r=48$).  
In this example, we observe that the contours produced by approximate inference methods are visually similar to those of exact inference methods. In addition, with increasing ranks, the contours produced by approximate inference methods approach those of the exact inference methods.

\begin{table}[h]
 \caption{Same as Table \ref{table:iact_elliptic}, but for PET and data set 1. The coordinate selection  is used in dimension reduction. Ranks of the subspaces are chosen to be $16$, $32$, and $48$.}\label{table:iact_pet}
\centering
\begin{tabular}{l@{\hspace{1\tabcolsep}}l| l@{\hspace{1\tabcolsep}}l@{\hspace{1\tabcolsep}}l @{\hspace{0.5\tabcolsep}}|@{\hspace{0.5\tabcolsep}} l@{\hspace{1\tabcolsep}}l@{\hspace{1\tabcolsep}}l | l @{\hspace{1\tabcolsep}}l}\hline
    && \multicolumn{2}{c}{IACT} & & \multicolumn{2}{c}{IACT} & & IACT \\ \cline{3-4}\cline{6-7}\cline{9-10}
    && \;OL & \;PM & \hspace{-1.2em}$\sqrt{\textrm{var}}[\log\widetilde{\mathcal{L}}_N^y]$\hspace{-0.3em} & \;OF & \;DA &  \;$\E[\beta]$ & HMALA & \;\;PCN \\ \hline
\multirow{3}{*}{\begin{sideways}$N=2$\end{sideways}} 
  & $r = 16$ & 33.2$\pm$1.7 & 85.1$\pm$2.7 & 1.54$\pm$0.02 & 33.9$\pm$1.1 & 214$\pm$44 & 0.18$\pm$0.06 & \multirow{6}{*}{95.9$\pm$3.3} & \multirow{6}{*}{387$\pm$79}\\
  & $r = 32$ & 40.0$\pm$1.8 & 54.1$\pm$3.1 & 0.61$\pm$.007 & 41.0$\pm$2.2 & 87.8$\pm$6.5 & 0.55$\pm$0.01 & \\
  & $r = 48$ & 45.3$\pm$1.2 & 49.4$\pm$2.6 & 0.45$\pm$.002 & 46.0$\pm$2.2 & 73.5$\pm$5.8 & 0.66$\pm$0.01 & \\ \cline{1-8}
\multirow{3}{*}{\begin{sideways}$N=5$\end{sideways}} 
  & $r = 16$ & 31.4$\pm$1.9 & 60.0$\pm$6.2 & 0.93$\pm$.006 & 31.8$\pm$1.4 & 220$\pm$65 & 0.22$\pm$0.04 & \\
  & $r = 32$ & 40.8$\pm$2.5 & 47.6$\pm$2.5 & 0.39$\pm$.004 & 42.8$\pm$2.4 & 88.0$\pm$6.8 & 0.56$\pm$0.01 & \\
  & $r = 48$ & 46.1$\pm$2.2 & 46.5$\pm$1.4 & 0.29$\pm$.001 & 46.3$\pm$1.9 & 69.5$\pm$4.0 & 0.67$\pm$0.01 & \\ \hline
\end{tabular}
\end{table}

We use the average IACTs of the density function, 
\(
\tau = \frac{1}{d}\sum_{i = 1}^d {\rm IACT}(f_i),
\) 
to measure the efficiency of various MCMC methods. The results are reported in Table \ref{table:iact_pet}. 
Here both PCN and H-MALA are implemented to sample the posterior in the transformed coordinate equipped with a Gaussian prior (cf. Section \ref{sec_normalization}).

For the approximate inference methods (OL and OF), the average IACTs consistently increase with the rank of the projectors, as the sampling performance of the Langevin proposal is expected to decay with underlying the parameter dimension. 
Both OL and OF produce significantly smaller IACTs compared with the full-dimensional PCN and H-MALA method. 
Compared to the OL method, the PM method, has a slightly higher IACTs in this example. This rather mild loss of performance (compared with the elliptic PDE example) is justified by the rather small values of $\widetilde{\mathcal{L}}_N^y$ (with $N=2,5$) in Table \ref{table:iact_pet}.
Compared to the OF method, the DA method, again produces the largest IACTs among all subspace inference methods.
However, the loss of performance here is not as severe as the elliptic PDE example, this is also justified by the improved second stage acceptance rates, $\E[\beta]$.

Overall, approximate inference methods have better statistical performance compared to other methods in this example and can obtain reasonably accurate results as shown in Figures \ref{fig:pet_KLD} and Figure \ref{fig:contour_pet}. With improved approximation errors, the exact inference methods also produces Markov chains with better mixing. Among all the exact inference methods, PM produces significantly smaller IACTs compared with other methods. 

\subsection{Numerical results using prior normalization}

\begin{figure}[h]
 \centering
 \includegraphics[trim=2.5cm 0cm 2cm 0cm, width = 0.8\textwidth]{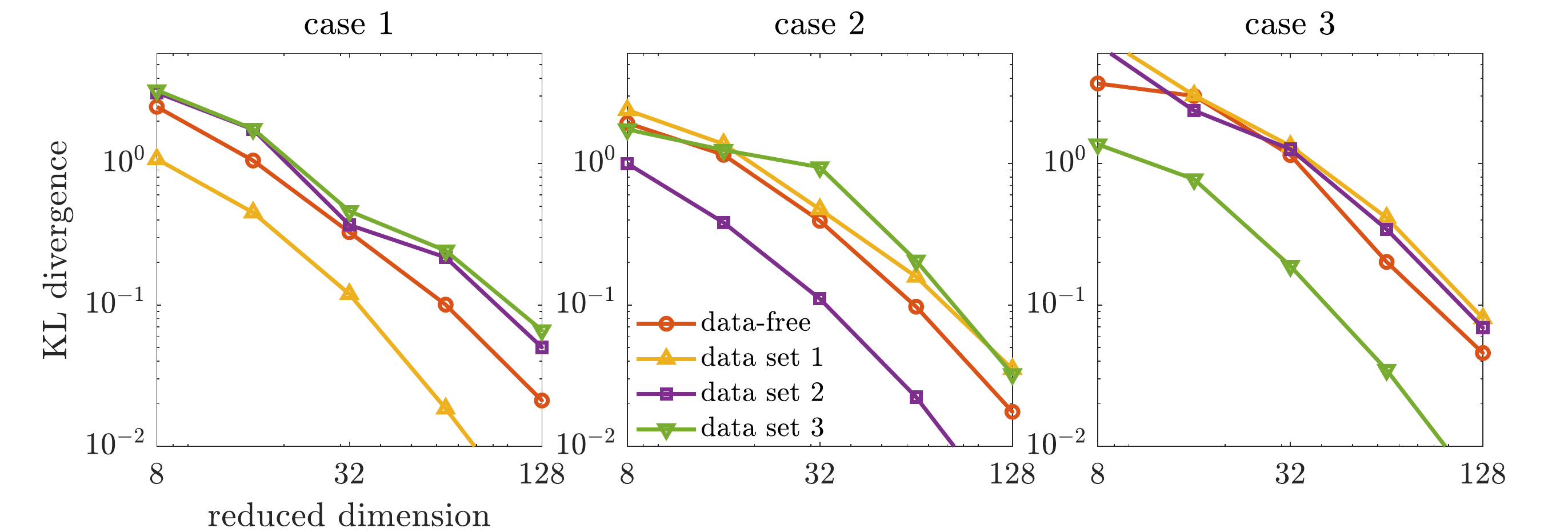}\\
 \includegraphics[trim=2.5cm 0cm 2cm 0cm, width = 0.8\textwidth]{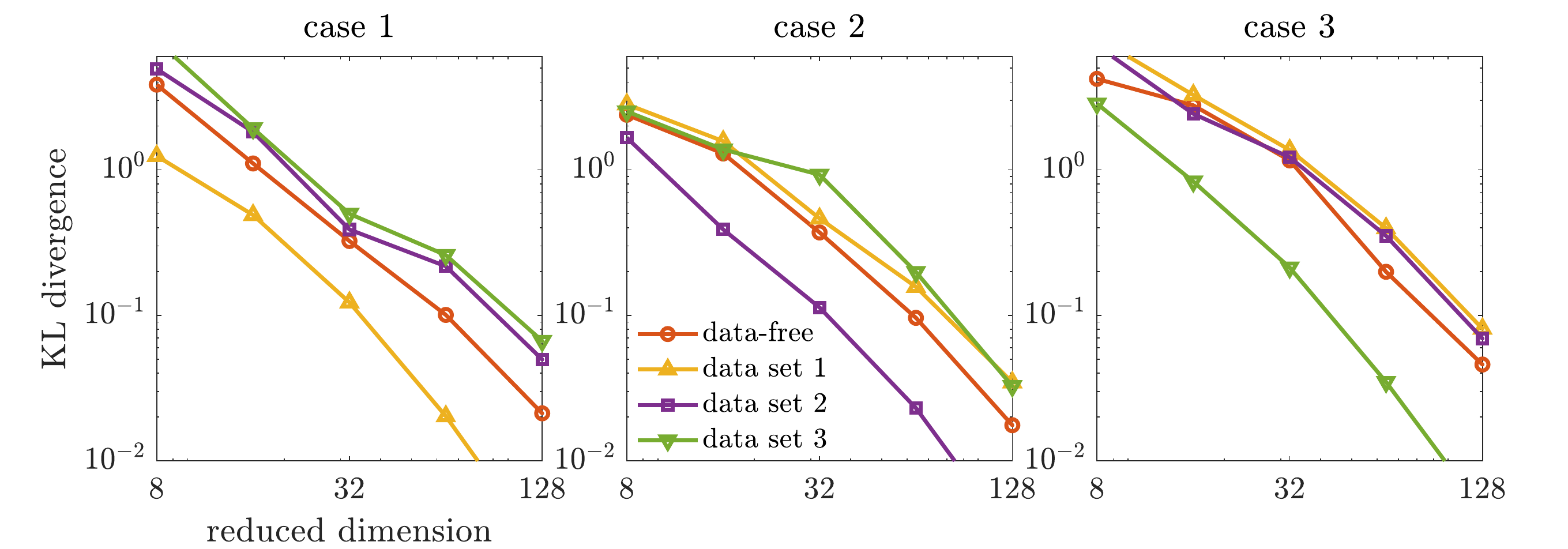}
 \caption{Same as Figure \ref{fig:elliptic_KLD}, but for PET. The prior normalization is used in dimension reduction.}\label{fig:pet_KLD_r}
\end{figure}

\begin{figure}[h]
 \centering
 \includegraphics[trim=2.5cm 1cm 2cm 0cm, width = 0.53\textwidth]{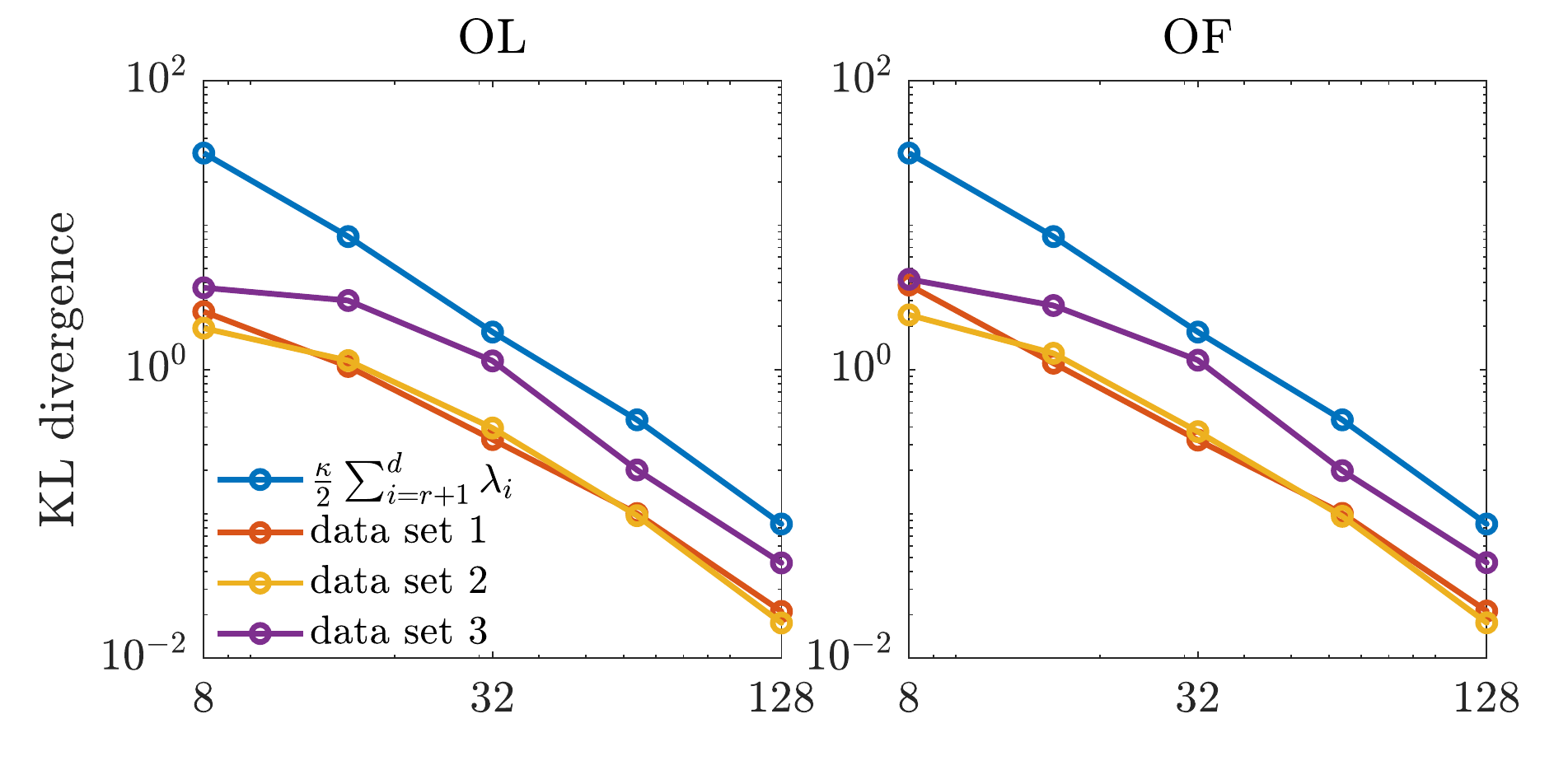}
 \caption{Same as Figure \ref{fig:ellptic_bound}, but for PET. The prior normalization is used in dimension reduction.}\label{fig:pet_bound_r}
\end{figure}

Then, we present the results obtained by applying the prior normalization method (cf. Section \ref{sec_normalization}). The KL divergences of the full posteriors from the approximated posterior densities  are shown in Figure \ref{fig:pet_KLD_r}. Here the result of prior-based dimension reduction is not presented, as the prior in the transformed space has an identity covariance matrix. 
We observe similar results as those obtained by the coordinate selection. We notice that the accuracy gaps between the data-free projectors and the data-dependent projectors (built using other data sets) are more significant compared with those obtained by the coordinate selection. 
The comparison of the errors of the approximate posterior densities (defined by the data-free projectors) with the bounds in \eqref{eq:KLBoundExpectation} and \eqref{eq:KLBoundExpectation_BIS} are provided in Figure \ref{fig:pet_bound_r}. Here we have $\kappa = 1$ because the transformed coordinate is endowed with a Gaussian prior. We observe that the errors of the approximate posterior densities follow the same trend as their corresponding bounds.
The IACTs of various MCMC methods are reported in Table \ref{table:iact_pet_r}. Again, the efficiency of subspace MCMC methods defined by the prior normalization is very close to that defined by the coordinate selection. 
Overall, both the coordinate selection and the prior normalization can be applied in this example to obtain accurate reduced-dimensional posterior approximations and derive efficient subspace MCMC methods. 

\begin{table}[h]
 \caption{Same as Table \ref{table:iact_elliptic}, but for PET and data set 1. The prior normalization is used in dimension reduction. Ranks of the subspaces are chosen to be $16$, $32$, and $48$.}\label{table:iact_pet_r}
\centering
\begin{tabular}{l@{\hspace{1\tabcolsep}}l| l@{\hspace{1\tabcolsep}}l@{\hspace{1\tabcolsep}}l @{\hspace{0.5\tabcolsep}}|@{\hspace{0.5\tabcolsep}} l@{\hspace{1\tabcolsep}}l@{\hspace{1\tabcolsep}}l | l @{\hspace{1\tabcolsep}}l}\hline
    && \multicolumn{2}{c}{IACT} & & \multicolumn{2}{c}{IACT} & & IACT \\ \cline{3-4}\cline{6-7}\cline{9-10}
    && \;OL & \;PM & \hspace{-1.2em}$\sqrt{\textrm{var}}[\log\widetilde{\mathcal{L}}_N^y]$\hspace{-0.3em} & \;OF & \;DA &  \;$\E[\beta]$ & HMALA & \;\;PCN \\ \hline
\multirow{3}{*}{\begin{sideways}$N=2$\end{sideways}} 
  & $r = 16$ & 35.8$\pm$1.7 & 81.4$\pm$6.0 & 1.48$\pm$0.02 & 33.5$\pm$1.8 & 168$\pm$23 & 0.25$\pm$0.02 & \multirow{6}{*}{95.9$\pm$3.3} & \multirow{6}{*}{387$\pm$79}\\
  & $r = 32$ & 42.8$\pm$2.0 & 55.1$\pm$2.9  & 0.64$\pm$.006 & 41.2$\pm$1.6 & 86.3$\pm$6.2 & 0.55$\pm$0.01 & \\
  & $r = 48$ & 45.0$\pm$2.4 & 51.8$\pm$2.0 & 0.46$\pm$.005 & 44.3$\pm$2.2 & 74.6$\pm$7.4 & 0.65$\pm$0.01 & \\ \cline{1-8}
\multirow{3}{*}{\begin{sideways}$N=5$\end{sideways}} 
  & $r = 16$ & 35.1$\pm$1.9 & 54.1$\pm$4.1 & 0.88$\pm$0.01 & 32.8$\pm$2.8 & 151$\pm$21 & 0.26$\pm$0.02 & \\
  & $r = 32$ & 45.0$\pm$1.7 & 49.0$\pm$1.9 & 0.41$\pm$.003 & 42.0$\pm$2.6 & 83.1$\pm$5.1 & 0.55$\pm$0.01 & \\
  & $r = 48$ & 45.9$\pm$2.9 & 46.3$\pm$2.2 & 0.29$\pm$.003 & 44.4$\pm$0.8 & 70.6$\pm$3.7 & 0.66$\pm$0.01 & \\ \hline
\end{tabular}
\end{table}

\section{Conclusion}

We present a new data-free strategy for reducing the dimensionality of large-scale statistical inverse problems. 
Compared to existing gradient-based dimension reduction technique, this new approach identifies the computationally costly subspace construction in an offline phase.
Our data-free dimension reduction is certified in the sense that its development is directly guided by factorizable posterior approximations and associated error bounds. 
The factorizable posterior approximations naturally offer dimension robust sampling methods for exploring the approximate posterior densities. 
More interestingly, by adding minor modifications to those approximate inference algorithms, we further develop exact inference methods  using the pseudo-marginal approach and the delayed acceptance approach. 
The resulting exact inference methods also scale well with parameter dimensionality, as the backbone of those methods is based on the dimension robust approximate inference methods.
We also demonstrate the efficiency of our data-free dimensional reduction and various inference methods on two inverse problems involving a two-dimensional elliptic PDE with a Gaussian process prior and a PET problem with Poisson data and a Besov-$\mathcal{B}^2_{11}$ prior. 

\ack{
T. Cui and O. Zahm would like to acknowledge support from the INRIA associate team UNQUESTIONABLE. T. Cui also acknowledges support from the Australian Research Council.}

\appendix

\section{Proof of Proposition \ref{prop:KLBoundExpectation_FORWARDMODEL}}
\label{proof:KLBoundExpectation_FORWARDMODEL}

 Recall $\pi_\text{pos}^y(x) = \frac{\mathcal{L}^y(x)  \pi_\text{pr}(x)}{\pi_\text{data}(y)}$ and $\widehat\pi_\text{pos}^y(x) = \frac{\widehat{\mathcal{L}}^y(x)  \pi_\text{pr}(x)}{\widehat\pi_\text{data}(y)}$.
 By definition of $\mathcal{L}^y(x)$ and $\widehat{\mathcal{L}}^y(x)$ we have
 \begin{align}
  \Dkl( &\pi_\text{pos}^y || \widehat\pi_\text{pos}^y )
  = \int_{\R^d}  \log \left(\frac{\pi_\text{pos}^y(x)}{\widehat\pi_\text{pos}^y(x)} \right) \pi_\text{pos}^y(x) \d x \nonumber\\
  &= \log\frac{\widehat\pi_\text{data}(y)}{\pi_\text{data}(y)} + \int_{\R^d} \left(  -\frac{1}{2}\| G(x)-y \|_{\Sigma_\text{obs}^{-1}}^2 + \frac{1}{2}\| G^\ast(x)-y \|_{\Sigma_\text{obs}^{-1}}^2  \right) \pi_\text{pos}^y(x) \d x \nonumber\\
  &= \log\frac{\widehat\pi_\text{data}(y)}{\pi_\text{data}(y)} + \int_{\R^d} \left( \frac{1}{2} \| e(x)\|_{\Sigma_\text{obs}^{-1}}^2 - e(x)^\top \Sigma_\text{obs}^{-1}(G(x)-y)   \right) \pi_\text{pos}^y(x) \d x \label{eq:tmp68721}
 \end{align}
 where $e(x)=G(x)-G^\ast(x)$ is independent on $y$.
 Next we replace $y$ by $Y\sim\pi_\text{data}$ and we take the expectation over $Y$.
 The first term in the above expression becomes
 \begin{equation}\label{eq:tmp7861}
  \E\left[\log\frac{\widehat\pi_\text{data}(Y)}{\pi_\text{data}(Y)}\right]
  = \int_{\R^m} \log\left(\frac{\widehat\pi_\text{data}(y)}{\pi_\text{data}(y)}\right) \pi_\text{data}(y)\d y = -\Dkl(\pi_\text{data}||\widehat\pi_\text{data}) .
 \end{equation}
 Next, by definition of $\pi_\text{pos}^y(x)$, we have
 \begin{align}
  \E\left[ \int_{\R^d}  \frac{1}{2} \| e(x)\|_{\Sigma_\text{obs}^{-1}}^2   \pi_\text{pos}^Y(x) \d x \right]
  &= \frac{1}{2}\int_{\R^d\times \R^m} \| e(x)\|_{\Sigma_\text{obs}^{-1}}^2   \pi_\text{pos}^y(x) \pi_\text{data}(y) \d x \d y \nonumber\\
  &= \frac{1}{2}\int_{\R^d\times \R^m} \| e(x)\|_{\Sigma_\text{obs}^{-1}}^2 \mathcal{L}^y(x) \pi_\text{pr}(x) \d x \d y \nonumber\\
  &= \frac{1}{2}\int_{\R^d} \| G(x)-\widehat G(x) \|_{\Sigma_\text{obs}^{-1}}^2 \pi_\text{pr}(x) \d x . \label{eq:tmp320999}
 \end{align}
 For the last equality we used the fact that $y\mapsto\mathcal{L}^y(x)$ is a pdf so that $\int_{\R^m} \mathcal{L}^y(x)\d y=1$.
 Using the same arguments, we have
 \begin{align}
  \E\bigg[ &\int_{\R^d} e(x)^\top \Sigma_\text{obs}^{-1}\big(G(x)-Y\big)  \pi_\text{pos}^Y(x) \d x \bigg]  \nonumber\\
  &= \int_{\R^d\times \R^m} e(x)^\top \Sigma_\text{obs}^{-1}\big(G(x)-y\big)  \mathcal{L}^y(x) \pi_\text{pr}(x)\d x \d y \nonumber\\
  &= \int_{\R^d } e(x)^\top \Sigma_\text{obs}^{-1}G(x) \pi_\text{pr}(x)\d x 
  - \int_{\R^d} e(x)^\top \Sigma_\text{obs}^{-1} \left(\int_{\R^m} y  \mathcal{L}^y(x) \d y  \right)\pi_\text{pr}(x)\d x  \nonumber\\
  &=0. \label{eq:tmp6824}
 \end{align}
 The last equality is obtained by noting that the expectation of the data knowing the parameter $x$ is $\int_{\R^m} y  \mathcal{L}^y(x) \d y = G(x)$. Combining \eqref{eq:tmp7861} \eqref{eq:tmp320999} and \eqref{eq:tmp6824}, we obtain 
 $$
  \E\left[ \Dkl( \pi_\text{pos}^Y || \widehat\pi_\text{pos}^Y ) \right] \overset{\eqref{eq:tmp68721}}{=} -\Dkl(\pi_\text{data}||\widehat\pi_\text{data}) + \frac{1}{2}\int_{\R^d} \| G(x)-\widehat G(x) \|_{\Sigma_\text{obs}^{-1}}^2 \pi_\text{pr}(x) \d x,
 $$
 which concludes the proof.

\section{Proof of Proposition \ref{prop:PseudoMarginalMCMC}}
\label{proof:prop:PseudoMarginalMCMC}

Consider a Metropolis-Hastings algorithm which targets the pdf $\pi_\text{tar}^{y,N}$ defined by \eqref{eq:piTar} using the following proposal density
\begin{equation}\label{eq:ProposalOnProductSpace}
 q\left(x_r^\dagger,\{x_\perp^{\dagger(i)}\}_{i=1}^N \Big |x_r,\{x_\perp^{(i)}\}_{i=1}^N \right)= q(x_r^\dagger|x_r)\prod_{i=1}^N \pi_\text{pr}(x_\perp^{\dagger(i)}|x_r^\dagger),
\end{equation}
where $q(x_r^\dagger|x_r)$ is the same proposal density as the one used at step 1 of Algorithm \eqref{alg:PseudoMarginalMCMC}.
The acceptance probability of this Metropolis-Hastings algorithm is given by
\begin{align*}
 \alpha\left(x_r^\dagger,\{x_\perp^{\dagger(i)}\}_{i=1}^N \Big |x_r,\{x_\perp^{(i)}\}_{i=1}^N \right) 
 &= \min\left\{1, \frac{\pi_\text{tar}^{y,N}(x_r^\dagger,\{x_\perp^{\dagger(i)}\}_{i=1}^N) ~
 q\left(x_r,\{x_\perp^{(i)}\}_{i=1}^N \Big |x_r^\dagger,\{x_\perp^{\dagger(i)}\}_{i=1}^N \right)}{\pi_\text{tar}^{y,N}(x_r,\{x_\perp^{(i)}\}_{i=1}^N)~ q\left(x_r^\dagger,\{x_\perp^{\dagger(i)}\}_{i=1}^N \Big |x_r,\{x_\perp^{(i)}\}_{i=1}^N \right)}\right\} \\
 &= \min\left\{1, \frac{\pi_\text{pr}(x_r^\dagger)\left(\sum_{i=1}^N \mathcal{L}^y(x_r^\dagger + x_\perp^{\dagger(i)}) \right)  q(x_r|x_r^\dagger)}{\pi_\text{pr}(x_r)\left(\sum_{i=1}^N \mathcal{L}^y(x_r + x_\perp^{(i)}) \right) q(x_r^\dagger|x_r)}\right\} ,
\end{align*}
which is precisely $\widehat\alpha_N(x_r^\dagger|x_r)$ defined in \eqref{eq:alphaHat}. Note that the first two steps of Algorithm \ref{alg:PseudoMarginalMCMC} consists of drawing a sample $(x_r^\dagger,\{x_\perp^{\dagger(i)}\}_{i=1}^N)$ from the proposal \eqref{eq:ProposalOnProductSpace}. This way, Algorithm \ref{alg:PseudoMarginalMCMC} can be interpreted as a MCMC algorithm which targets $\pi_\text{tar}^{y,N}$. 
It remains to show that the marginal distribution $\pi_\text{tar}^{y,N}(x_r)$ is the marginal posterior $\pi_\text{pos}^y(x_r)$. We can write
\begin{align*}
 \pi_\text{tar}^{y,N}(x_r) 
 &= \int_{\text{Ker}(P_r)^N} \pi_\text{tar}^{y,N}(x_r,x_\perp^{(1)},\ldots,x_\perp^{(N)}) \d x_\perp^{(1)} \ldots\d x_\perp^{(N)} \\
 &\propto \pi_\text{pr}(x_r)\sum_{i=1}^N \int_{\text{Ker}(P_r)} \mathcal{L}^y(x_r + x_\perp^{(i)}) \pi_\text{pr}(x_\perp^{(i)}|x_r)\d x_\perp^{(i)} \\
 &\propto \int_{\text{Ker}(P_r)} \mathcal{L}^y(x_r + x_\perp^{(i)}) \pi_\text{pr}(x_r+x_\perp^{(i)})\d x_\perp \propto \pi_\text{pos}^y(x_r)
\end{align*}
which concludes the proof.

\section{Proof of Proposition \ref{prop:MCMC_for_exact_inference}}
\label{proof:prop:MCMC_for_exact_inference}

Recall that $\{(X_r^{(j)} , \{X_\perp^{(j,i)}\}_{i=1}^N )\}_{j\geq1}$ admits $\pi_\text{tar}^{y,N}$ \eqref{eq:piTar} as the invariant density, see Proposition \ref{prop:PseudoMarginalMCMC}. It remains to prove that $\{X^{(j)}\}_{j\geq 1}$ admits $\pi_\text{pos}^y(x)$ as the invariant density. For a given state  $(X_r^{(j)} , \{X_\perp^{(j,i)}\} = (x_r,\{x_\perp^{(i)}\}_{i=1}^N)$, we have $X^{(j)} = x_r+x_\perp$ where $x_\perp\in\{x_\perp^{(i)}\}_{i=1}^N$ is selected with respect to the probability
\begin{equation}\label{eq:x_perpStar_PROOF}
 \mathbb{P}\left(x_\perp = x_\perp^{(k)} \Big| x_r ,\{x_\perp^{(i)}\}_{i=1}^N  \right) = \frac{\mathcal{L}^y(x_r+x_\perp^{(k)})}{\sum_{i=1}^N \mathcal{L}^y(x_r+x_\perp^{(i)})},
  \quad 1\leq k\leq N.
\end{equation}
Thus, we need to prove that the pdf $\pi(x)$ where $x = x_r+x_\perp$ is the posterior density $\pi(x)=\pi_\text{pos}^y(x)$.
We can write
\begin{align*}
 \pi(x) 
 &=\pi(x_r,x_\perp) \\
 &= \int_{\text{Ker}(P_r)^N}\pi(x_r,\{x_\perp^{(i)}\}_{i=1}^N,x_\perp) \,\d x_\perp^{(1)}\ldots \d x_\perp^{(N)} \\
 &= \int_{\text{Ker}(P_r)^N} \pi\left(x_\perp \Big| x_r,\{x_\perp^{(i)}\}_{i=1}^N\right)\pi_\text{tar}^{y,N}\left( x_r,\{x_\perp^{(i)}\}_{i=1}^N\right) \,\d x_\perp^{(1)}\ldots \d x_\perp^{(N)},
\end{align*}
where $\pi(x_\perp | x_r,\{x_\perp^{(i)}\}_{i=1}^N )$ is the pdf of $x_\perp$ conditioned on $(x_r,\{x_\perp^{(i)}\}_{i=1}^N)$. By construction we have
\begin{equation}\label{eq:tmp243678}
 \pi\left(x_\perp \Big| x_r,\{x_\perp^{(i)}\}_{i=1}^N \right) 
 \overset{\eqref{eq:x_perpStar_PROOF}}{=} \frac{\sum_{k=1}^N \delta_{x_\perp^{(k)}}(x_\perp) \mathcal{L}^y(x_r+x_\perp^{(k)}) }{\sum_{i=1}^N \mathcal{L}^y(x_r+x_\perp^{(i)})},
\end{equation}
where $\delta_{x_\perp^{(k)}}$ denotes the Dirac mass function at point $x_\perp^{(k)}$. We can write
\begin{align*}
 \pi(x) 
 &=\int_{\text{Ker}(P_r)^N}  \frac{\sum_{k=1}^N \delta_{x_\perp^{(k)}}(x_\perp) \mathcal{L}^y(x_r+x_\perp^{(k)}) }{\sum_{i=1}^N \mathcal{L}^y(x_r+x_\perp^{(i)})} \pi_\text{tar}^{y,N}\left( x_r,\{x_\perp^{(i)}\}_{i=1}^N\right) \,\d x_\perp^{(1)}\ldots \d x_\perp^{(N)} \\
 &\overset{\eqref{eq:piTar}}{\propto} \sum_{k=1}^N\int_{\text{Ker}(P_r)^N} \delta_{x_\perp^{(k)}}(x_\perp) \mathcal{L}^y(x_r+x_\perp^{(k)})  \pi_\text{pr}(x_r) \prod_{i=1}^N \pi_\text{pr}(x_\perp^{(i)}|x_r) \,\d x_\perp^{(1)}\ldots \d x_\perp^{(N)} \\
 &\propto \sum_{k=1}\int_{\text{Ker}(P_r)} \delta_{x_\perp^{(k)}}(x_\perp) \mathcal{L}^y(x_r+x_\perp^{(k)})  \pi_\text{pr}(x_r) \pi_\text{pr}(x_\perp^{(k)}|x_r) \,\d x_\perp^{(k)} \\
 &\propto \sum_{k=1} \mathcal{L}^y(x_r+x_\perp)  \pi_\text{pr}(x_r) \pi_\text{pr}(x_\perp|x_r) 
 \propto \pi_\text{pos}^y(x_r+x_\perp),
\end{align*}
which concludes the proof. 

\section{Proof of Proposition \ref{prop:DAMCMC}}
\label{proof:prop:DAMCMC}
To show the result of Proposition \ref{prop:DAMCMC}, we first interpret the first stage acceptance/rejection and the conditional prior sampling $\pi_{\rm pr}(x_\perp^\dagger | x_r^\dagger)$ as a joint proposal acting in the full parameter space ${\rm Im}(P_r) \oplus {\rm Ker}(P_r)$. 
The proposal $q(\cdot, | x_r)$ and the acceptance probability $\alpha(x_r^\dagger|x_r)$ defines an effective proposal distribution 
\[
\bar q(x_r^\dagger|x_r) = q(x_r^\dagger|x_r )\alpha(x_r^\dagger|x_r) + \Big[1 - \int q(x_r^\dagger|x_r )\alpha(x_r^\prime|x_r) dx_r^\prime\Big]\delta_{x_r}(x_r^\dagger),
\]
where $\delta_{x_r}(\cdot)$ denotes the Dirac delta and the term in the bracket represents the probability of a proposal candidate being rejected. Then, we can define a joint proposal distribution
\[
Q(x_r^\dagger, x_\perp^\dagger | x_r, x_\perp) := Q(x_r^\dagger, x_\perp^\dagger | x_r) = \pi_{\rm pr}(x_\perp^\dagger | x_r^\dagger)\,\bar q(x_r^\dagger | x_r),
\]
for the MH to sample the full posterior.

Following the exactly same derivation in \cite{christen2005markov}, one can show that accepting $(x_r^\dagger, x_\perp^\dagger) \sim Q(\cdot, \cdot | x_r, x_\perp)$ with the probability 
\begin{align*}
\beta(x_r^\dagger,x_\perp^\dagger | x_r,x_\perp) & = \min\left[ 1, \frac{\pi^{y}_{\rm pos}(x_r^\dagger+x_\perp^\dagger) \,Q(x_r, x_\perp | x_r^\dagger, x_\perp^\dagger)} {\pi^{y}_{\rm pos}(x_r+ x_\perp)\,Q(x_r^\dagger, x_\perp^\dagger | x_r,x_\perp)} \right] 
\end{align*}
defines a Markov transition kernel with the full posterior $\pi^{y}_{\rm pos}(x_r+ x_\perp)$ as its invariant distribution. Since the above acceptance probability is only used in the case where the first stage proposal candidate $x_r^\dagger$ is accepted, i.e., $x_r^\dagger \neq x_r$, we do not need to consider the Dirac delta term. This way, the above acceptance probability can be simplified as
\begin{align*}
\beta(x_r^\dagger,x_\perp^\dagger | x_r,x_\perp) = \min\left[ 1, \frac{\pi^{y}_{\rm pos}(x_r^\dagger+x_\perp^\dagger) \,\pi_{\rm pr}(x_\perp | x_r)\,q(x_r|x_r^\dagger)\,\alpha(x_r|x_r^\dagger)} {\pi^{y}_{\rm pos}(x_r+ x_\perp)\,\pi_{\rm pr}(x_\perp^\dagger | x_r^\dagger)\,q(x_r^\dagger|x_r )\alpha(x_r^\dagger|x_r)} \right].
\end{align*}
Substituting the identities
\[
\frac{\alpha(x_r|x_r^\dagger)} {\alpha(x_r^\dagger|x_r)}  = \frac{\widetilde{\mathcal{L}}_N^y( x_r ) \pi_\text{pr}(x_r)\,q(x_r^\dagger | x_r)}{\widetilde{\mathcal{L}}_N^y( x_r^\dagger ) \pi_\text{pr}(x_r^\dagger)\,q(x_r | x_r^\dagger)}, 
\]
and
\[
\frac{\pi^{y}_{\rm pos}(x_r^\dagger+x_\perp^\dagger) \,\pi_{\rm pr}(x_\perp | x_r)} {\pi^{y}_{\rm pos}(x_r+ x_\perp)\,\pi_{\rm pr}(x_\perp^\dagger | x_r^\dagger)}  = \frac{\mathcal{L}^y( x_r^\dagger + x_\perp^\dagger )\,\pi_{\rm pr}( x_r^\dagger)} {\mathcal{L}^y( x_r + x_\perp )\,\pi_{\rm pr}( x_r)},
\]
into the above equation, we obtain 
\[
\beta(x_r^\dagger,x_\perp^\dagger | x_r,x_\perp) = \min\left[ 1, \frac{\mathcal{L}^y(x_r^\dagger+ x_\perp^\dagger) \,\widetilde{\mathcal{L}}_N^y(x_r)} {\mathcal{L}^y( x_r + x_\perp )\,\widetilde{\mathcal{L}}_N^y(x_r^\dagger)}\right],
\]
which is identical to the second stage acceptance probability in \eqref{eq:DA_acc2}. Thus, the result follows. 

\section{Cumulative density function of \texorpdfstring{$p(x) \propto \exp( - \gamma |x|^p)$}{sth}}
\label{sec:cdf_Besov}
Given the pdf
\(
p(x) = \frac{1}{c_{\gamma,p}} \exp( - \gamma |x|^p),
\)
$x\in\R$, we want to find its normalizing constant $c_{\gamma,p}$ and cdf.
Using symmetry, the normalizing constant takes the form
\[
c_{\gamma,p} = 2 \int_{0}^{\infty} \exp( - \gamma x^p) dx,
\]
and the cdf can be expressed as
\[
\Phi(x) = \left\{ \begin{array}{ll} \displaystyle \frac12 + \frac{1}{c_{\gamma,p}}\int_{0}^{x} \;\;\exp( - \gamma t^p) dt & x\geq 0 \vspace{6pt }\\  \displaystyle \frac12 - \frac{1}{c_{\gamma,p}}\int_{0}^{-x} \exp( - \gamma t^p) dt & x <0 \end{array} \right..
\]
We first introduce the change-of-variable $z = \gamma t^p$ so that 
\[
t = \bigg(\frac{z}{\gamma}\bigg)^{\frac1p} \quad {\rm and} \quad \frac{\d t}{\d z} = p^{-1} \gamma^{-\frac1{p}} \, z^{\frac1p - 1}.
\]
This yields
\[
\int_{0}^{x} \exp( - \gamma t^p) dt = p^{-1} \gamma^{-\frac1{p}} \int_{0}^{\gamma x^p}  z^{\frac1p - 1} \exp( - z) dz := p^{-1} \gamma^{-\frac1{p}} \Gamma_\text{lower}(p^{-1}, \gamma x^p) ,
\]
where $\Gamma_\text{lower}(\cdot, \cdot)$ is the lower incomplete gamma function. Following a similar derivation, we obtain 
\(
c_{\gamma,p} = 2 \,p^{-1} \gamma^{-\frac1{p}} \,\Gamma(p^{-1}),
\)
where $\Gamma(\cdot)$ is the Gamma function. This way, we have the cdf
\[
\Phi(x) = \frac12 + \frac{{\rm sign}(x)}{2 \, \Gamma(p^{-1})} \Gamma_\text{lower}\Big(p^{-1}, \gamma \,\big({\rm sign}(x)\, x\big)^p \Big).
\]
There are two notable special cases. The Gaussian distribution $\mathcal{N}(0, \sigma^2)$ can be specified using $\gamma = (2\sigma^2)^{-1}$ and $p = 2$, in which the cdf can be equivalently expressed using the error function. 
The Laplace distribution can be specified using $p = 1$, so that the cdf yields a simpler (but equivalent) expression in the form of
\[
\Phi(x) = \frac12 + \frac{{\rm sign}(x)}{2} \Big( 1 - \exp\big( - \gamma \,{\rm sign}(x)\,x\big) \Big).
\]

\section{Derivation of Fisher information matrices}
\label{sec:fisher}

Here we derive the Fisher information matrix for the Poisson likelihood case. Recall the Fisher information matrix  
\begin{equation}
 \mathcal{I}(x) = \int_{\R^m}  \big( \nabla \log \mathcal{L}^y(x) )( \nabla \log \mathcal{L}^y(x) \big)^\top \,\mathcal{L}^y(x)\, \d y .
\end{equation}
Defining the predata $\eta = G(x)$, we can express the gradient of the likelihood function as
\[
\nabla_x \log \mathcal{L}^y(x) = \nabla G(x)^\top \nabla_\eta \log \mathcal{L}^y(\eta).
\]
where
\[
\mathcal{L}^y(\eta) = \prod_{i = 1}^{m} \frac{\eta_i^{y_i}\, \exp(- \eta_i) }{y_i !}, \quad {\rm subject\;to} \quad \eta = G(x).
\]
This way, the Fisher information matrix can be rewritten as
\begin{equation}
 \mathcal{I}(x) = \nabla G(x)^\top \bigg( \int_{\R^m}  \big( \nabla \log \mathcal{L}^y(\eta) )( \nabla \log \mathcal{L}^y(\eta) \big)^\top \,\mathcal{L}^y(\eta)\, \d y \bigg) \nabla G(x), 
\end{equation}
subject to $\eta = G(x)$. The term in the brackets of the above equation is the Fisher information matrix of the Poisson distribution, which is a diagonal matrix 
\[
\bigg( \int_{\R^m}  \big( \nabla \log \mathcal{L}^y(\eta) )( \nabla \log \mathcal{L}^y(\eta) \big)^\top \,\mathcal{L}^y(\eta)\, \d y \bigg)_{ii} = \frac{1}{\eta_i}. 
\]
Thus, the Fisher information matrix w.r.t. $x$ is
\begin{align}
\mathcal{I}(x) = \nabla G(x)^\top M(x) \nabla G(x),
\end{align}
where $M(x)$ is a diagonal matrix with $M_{ii}(x) = G_i(x)^{-1}$ along its diagonal.

{
\section*{References}
\bibliographystyle{plain}
\bibliography{references} 
}

\end{document}